\documentclass[aps,reprint,pre,showpacs,superscriptaddress]{revtex4-2}
\usepackage{amsmath}
\usepackage{amssymb}
\usepackage{graphicx}
\usepackage{xcolor}
\usepackage{bm}
\usepackage{mathtools}
\usepackage{MnSymbol}
\usepackage{array}
\usepackage{multirow}
\usepackage{dcolumn} 
\usepackage{siunitx}
\definecolor{linkcolor}{rgb}{0.56640625,0.3359375,0.42578125} 

\usepackage[pdftex,colorlinks=true,
pdfstartview=FitV,
linkcolor= linkcolor,
citecolor= linkcolor,
urlcolor= linkcolor,
hyperindex=true,
hyperfigures=false]
{hyperref}

\newcommand{\modellong}{Suspension Hébraud-Lequeux model}
\newcommand{\modelshort}{SHL}
\newcommand{\levymodelshort}{L\'evy-SHL}
\newcommand{\mediatedROMinf}{MROM$_{\infty}$}

\begin{document}

\title{Random organization criticality with long-range hydrodynamic interactions}

\author{Tristan Jocteur}
\affiliation{Univ. Grenoble-Alpes, CNRS, LIPhy, 38000 Grenoble, France}
\author{Cesare Nardini}
\affiliation{Service de Physique de l’État Condensé, CNRS UMR 3680, CEA-Saclay, 91191 Gif-sur-Yvette, France}
\author{Eric Bertin}
\author{Romain Mari}
\affiliation{Univ. Grenoble-Alpes, CNRS, LIPhy, 38000 Grenoble, France}

\begin{abstract}
Driven soft athermal systems may display a reversible-irreversible transition between an absorbing, arrested state and an active phase where a steady-state dynamics sets in. 
A paradigmatic example consists in cyclically sheared suspensions under stroboscopic observation, for which in absence of contacts during a shear cycle particle trajectories are reversible and the stroboscopic dynamics is frozen, while contacts lead to diffusive stroboscopic motion. 
The Random Organization Model (ROM), which is a minimal model of the transition, shows a transition which falls into the Conserved Directed Percolation (CDP) universality class. 
However, the ROM ignores hydrodynamic interactions between suspended particles, which make contacts a source of long-range mechanical noise that in turn can create new contacts.
Here, we generalize the ROM to include long-range interactions decaying like inverse power laws of the distance.
Critical properties  continuously depend on the decay exponent when it is smaller than the space dimension.
Upon increasing the interaction range, the transition turns  convex (that is, with an order parameter exponent $\beta>1$), fluctuations turn from diverging to vanishing, and hyperuniformity at the transition disappears.
We rationalize this critical behavior using a local mean-field model describing how particle contacts are created via mechanical noise, showing that diffusive motion induced by long-range interactions becomes dominant for slowly-decaying interactions.
\end{abstract}

\maketitle

\section{Introduction}

For soft athermal materials, such as foams, granular materials or emulsions, 
external forces are the only source of motion at the microscopic scale, 
so that dynamic arrest naturally occurs in the absence of driving.
Yet, it is common that even in the presence of external driving, the dynamics becomes arrested, in states called ``absorbing states''.
For instance, yield stress fluids jam into a static state in mechanical equilibrium when subject to a stress below the yield value~\cite{bonn2017yield}, systems of active particles can freeze if activity is conditional due to e.g. some sort of quorum sensing~\cite{leiHydrodynamicsRandomorganizingHyperuniform2019a,leiNonequilibriumStronglyHyperuniform2019,golestanianBoseEinsteinlikeCondensationScalar2019,LiuActivityPNAS2021,lefrancQuorumSensingAbsorbing2025}. 

Soft materials under cyclic shear are a paradigmatic class of systems showing arrested dynamics with absorbing states. 
When subject to an oscillatory shear with small enough driving amplitude, after a finite number of shear cycles the system reaches a limit cycle where the microscopic configuration keeps visiting the same trajectory over and over. 
As a consequence, when observed stroboscopically (once per cycle), the dynamics is arrested, with strictly frozen microscopic configuration.
This only occurs below a critical driving amplitude though, as if one exceeds this threshold, the stroboscopic dynamics is diffusive even after arbitrary large number of cycles. 
Initially discovered in non-Brownian suspensions~\cite{pineChaosThresholdIrreversibility2005,corteRandomOrganizationPeriodically2008,corteSelfOrganizedCriticalitySheared2009,guastoHydrodynamicIrreversibilityParticle2010,geRheologyPeriodicallySheared2022}, this transition, coined Reversible-Irreversible Transition (RIT), has since been observed in virtually all athermal complex fluids~\cite{reichhardtReversibleIrreversibleTransitions2023}, like dry granular materials~\cite{royerPreciselyCyclicSand2015}, microemulsions~\cite{jeanneretGeometricallyProtectedReversibility2014,weijsEmergentHyperuniformityPeriodically2015}, foams~\cite{mukherjiStrengthMechanicalMemories2019} or soft glasses~\cite{fioccoOscillatoryAthermalQuasistatic2013,keimMechanicalMicroscopicProperties2014,himanagamanasaExperimentalSignaturesNonequilibrium2014,regevReversibilityCriticalityAmorphous2015,ederaYieldingMicroscopeMultiscale2024}.

There is a minimal set of recurring ingredients shared by systems showing RIT.
In all cases, irreversibility in a shearing cycle occurs through individual, separable events (as opposed to the irreversibility of a chaotic dynamical system for instance).
For dilute suspensions, these are particle contacts occurring during the cycles, which break the time-reversibility otherwise ensured by Stokes hydrodynamics~\cite{corteRandomOrganizationPeriodically2008}.
For dense, jammed systems, 
irreversibility is a consequence of plastic events~\cite{princenRheologyFoamsHighly1983,picardSlowFlowsYield2005,nessAbsorbingStateTransitionsGranular2020}.
Cycle reversibility is then achieved by annealing out irreversible events through repeated shearing cycles, to end up for instance in a contact-less cycle in the case of suspensions or a purely elastic cycle in the case of jammed systems.
(We ignore in this discussion the possibility of limit cycles of repeated plasticity~\cite{munganCyclicAnnealingIterated2019,munganNetworksHierarchiesHow2019}, which might be a finite-size effect~\cite{kumarMappingOutGlassy2022,munganNetworksHierarchiesHow2019,liuFateShearoscillatedAmorphous2022}.)

Being the first and arguably the simplest realization of RIT, cyclically sheared suspensions are a key platform to explore this transition, and are more generally a paradigm for dynamical arrest in soft materials.
The Random Organization Model (ROM)~\cite{corteRandomOrganizationPeriodically2008,hexnerHyperuniformityCriticalAbsorbing2015,hexnerNoiseDiffusionHyperuniformity2017,tjhungHyperuniformDensityFluctuations2015,schrenkCommunicationEvidenceNonergodicity2015} was introduced as a minimal model for the stroboscopic dynamics.
It was instrumental in allowing one to discuss the classification of the RIT in suspensions. 
Indeed, it is akin to a continuum version of lattice-based stochastic sandpile models~\cite{dharAbelianSandpileRelated1999}, such as the Manna model~\cite{mannaTwostateModelSelforganized1991}.
This led to a widespread expectation~\cite{corteRandomOrganizationPeriodically2008,menonUniversalityClassReversibleirreversible2009,nessAbsorbingStateTransitionsGranular2020} that  the suspension RIT belongs to the Conserved Directed Percolation universality class (CDP)~\cite{vespignani1998driving,vespignaniAbsorbingstatePhaseTransitions2000a,rossiUniversalityClassAbsorbing2000,henkelAbsorbingPhaseTransitions2008}.
Indeed, the ROM satisfies the properties conjectured to ensure that a model belongs to CDP: 
it is a stochastic model with an infinite number of absorbing states in which the order parameter (the density of active particles) is coupled to a non-diffusive conserved field (the total density of particles).
Simulations of the ROM tend to confirm this expectation~\cite{hexnerHyperuniformityCriticalAbsorbing2015,hexnerNoiseDiffusionHyperuniformity2017,mariAbsorbingPhaseTransitions2022} (see however~\cite{tjhungCriticalityCorrelatedDynamics2016}), and in particular it is found that for $\phi \to \phi_\mathrm{c}^+$ (where $\phi_\mathrm{c}$ is the critical packing fraction), the long-time density of active particles $\langle A\rangle$ scales as $A \sim (\phi-\phi_\mathrm{c})^\beta$ with $\beta\approx 0.64$ in spatial dimension $d=2$~\cite{mariAbsorbingPhaseTransitions2022} in good agreement with the exponents reported for the CDP class~\cite{lubeckUniversalScalingBehavior2004}.


However, assessing if the RIT does belong to CDP in actual experiments or simulations of cyclically-sheared suspensions proves challenging.
Initial claims of compatibility with CDP in cyclically sheared suspensions ~\cite{corteRandomOrganizationPeriodically2008} have been challenged in experiments on emulsions 
~\cite{jeanneretGeometricallyProtectedReversibility2014,weijsEmergentHyperuniformityPeriodically2015} or numerical simulations of dense frictionless disks~\cite{nagasawaClassificationReversibleIrreversible2019}.
Numerical simulations that allow more accuracy close to the transition, as they do not suffer from position tracking uncertainties, show a convex RIT, that is, with $\beta>1$~\cite{agrawalDenseSuspensionsRotary2024,metzgerIrreversibilityChaosRole2010,metzgerIrreversibilityChaosRole2013}\footnote{In the original articles~\cite{agrawalDenseSuspensionsRotary2024,metzgerIrreversibilityChaosRole2010,metzgerIrreversibilityChaosRole2013}, the diffusion coefficients as a function of strain amplitude are not plotted in a convention that allows to easily grasp the convex character of the RIT, but if one plots these data in a lin-lin representation, the convexity is striking.}.
This  discrepancy with the concave transition predicted by CDP suggests that something fundamental is missed in the ROM.

Some CDP-incompatible behaviors have been argued to stem from the presence of long-ranged hydrodynamic interactions~\cite{weijsEmergentHyperuniformityPeriodically2015}, falling as $1/r^\alpha$ with $r$ the distance between a particle and a contact in the suspension.
Contacts between ``active'' particles during a shear cycle are felt by every other particle, which results in irreversible trajectories for all of them
~\cite{weijsEmergentHyperuniformityPeriodically2015}.
In a recent work~\cite{mariAbsorbingPhaseTransitions2022}, some of us argued that this gives a diffusive dynamics to passive particles too, which 
creates a new channel for creation of activity through contacts of two passive particles.
A model based on the ROM with homogenized activity-induced diffusion of passive particles (i.e.  function of the average activity density in the system) indeed exhibits a convex transition with $\beta\approx 1.85$~\cite{mariAbsorbingPhaseTransitions2022}.
This homogenized diffusion however potentially washes out effects from actual spatialized hydrodynamic interactions.
Moreover, it is important to characterize the dependence of the RIT on the range of interactions, that is, on the value of $\alpha$, as in experiments this value can change, depending in particular on whether the system is a bulk ($\alpha=d-1$) or a confined ($\alpha=d$) suspension~\cite{diamantHydrodynamicInteractionConfined2009}.

In this article, we thus generalize the model introduced in~\cite{mariAbsorbingPhaseTransitions2022} to account for spatialized long-range hydrodynamic interactions, systematically varying the range exponent $\alpha$, in spatial dimensions $d=2$ and $d=3$. 
We measure the values of the critical exponents for the reversible-irreversible transition as a function of $\alpha$.
We show that the homogenized diffusion behavior is recovered for small values of $\alpha$, that  the critical exponents continuously evolve when $\alpha$ increases up to $\alpha \approx d$, and that for $\alpha\gtrsim d$ the transition falls into usual CDP.
This implies that RITs in bulk and confined systems do not belong to the same universality class and have markedly different behaviors: we expect the former to be convex transitions ($\beta>1$) whereas the latter are in the CDP class. 
Furthermore, small values of $\alpha$ that are associated with a convex transition also show other unusual features: activity fluctuations vanish at the transition, and hyperuniformity, which is a prominent feature of the CDP transition~\cite{hexnerHyperuniformityCriticalAbsorbing2015,hexnerEnhancedHyperuniformityRandom2017,hexnerNoiseDiffusionHyperuniformity2017,tjhungHyperuniformDensityFluctuations2015,tjhungCriticalityCorrelatedDynamics2016,schrenkCommunicationEvidenceNonergodicity2015,maHyperuniformityGeneralizedRandom2019}, is lost.
Hyperuniformity may however remain at larger $\alpha$ values where the transition is concave but does not belong to CDP.

\section{Model}

In order to model the critical behavior of cyclically sheared suspensions, we develop a model based on the Random Organization Model (ROM). In this section we first present the original ROM before introducing our model.

\subsection{Random Organization Model}

The ROM is a minimal model describing the stroboscopic dynamics of the suspension. 
We consider a $d$-dimentional cubic box with periodic boundary conditions, of linear extension $L$ and volume $L^d$.
Within this box, there are $N$ spheres of diameter $D$. 
We call $\bm{r}^t_i$ the position of particle $i$ at time $t$.
Particles can overlap, that is for a pair of particles $i$ and $j$, their distance $r^t_{ij} = |\bm{r}^t_i -\bm{r}^t_j|$ (accounting for periodic boundary conditions) is such that $r^t_{ij}<D$.
When this happens, particles $i$ and $j$ are considered active.
Active particles in the ROM would correspond to particles that make contact during a shear cycle in an actual experiment. 
The diameter $D$ is thus an effective diameter taking into account the underlying shear amplitude (which is not described explicitly in this stroboscopic model), and not the physical diameter of suspended particles.
Conversely, particles that are not involved in any overlap are considered passive. 
The dynamics of the ROM is in discretized time.
At time $t$, every active particle (say, $i$) is displaced by a random kick $\bm{\delta}_{a,i}$ taken within a distribution with typical kick size $\Delta_a$ (which in this article we pick as $\Delta_a = D$), such that
\begin{equation}
    \bm{r}^{t+1}_i = \bm{r}^{t}_i + \bm{\delta}_{a,i}\, .
\end{equation}
These kicks represent the displacements which result from contacts during the cycle in an actual suspension.
At the end of the time step, an active particle $i$ remains active if it is still in an overlap (that is, there is a particle $j$ such that $r^{t+1}_{ij} < D$) or otherwise becomes passive.
In turn, passive particles can become active if an active particle overlaps them after being given a random kick.
The dynamics runs as long as there are active particles. 
When all particles are passive, the system reaches an absorbing state: no particle moves, so that no overlap can occur, and therefore the dynamics is forever frozen.

Depending on the particle volume fraction $\phi = N v_d (D/2)^d/L^d$ (with $v_d$ the volume of a $d$-dimensional unit sphere), at long times two behaviors can be observed. 
If $\phi$ is smaller than some critical density $\phi_\mathrm{c}$, the system will always fall into an absorbing state with no overlaps. 
However if $\phi > \phi_\mathrm{c}$, the system will reach a stationary state for which the proportion of active particles $A$ will fluctuate around an average value $\langle A \rangle>0$. 
This model then exhibits an absorbing phase transition with control parameter $\phi$ and order parameter $\langle A \rangle$. 

\subsection{Mediated ROM}

\begin{figure}
    \centering
    \includegraphics[width=0.45\textwidth]{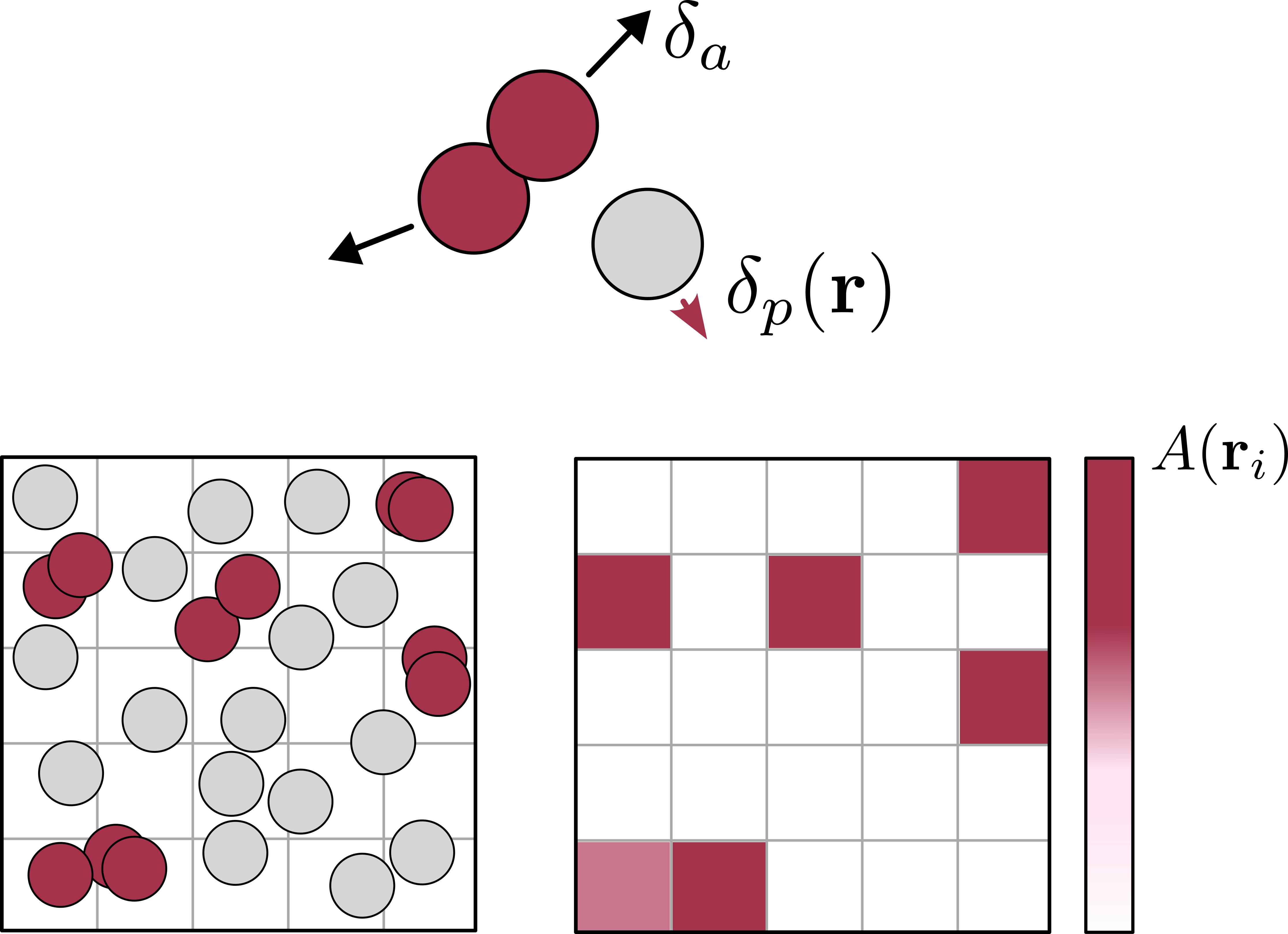}
    \caption{(a) Description of the rules for the mediated ROM. Active particles undergo a random kick $\delta_a$ of typical size $\Delta_a$ while passive particles are given a random kick $\delta_p(\bm{r})$ of typical size $\Delta_p(\bm{r})$ that depends on their position. (b) Determination of the spatialized activity field for the computation of passive particles steps.}
    \label{fig:rules}
\end{figure}

The ROM relies on a strong simplification: passive particles do not move. 
This corresponds to assuming that in an actual experiment particles that do not get in contact  during a shear cycle follow reversible trajectories. 
However, this is not the case unless all particles are contactless during the cycle, that is, unless the system is in an absorbing state.
As soon as there is one contact during the shear cycle, the contact force is balanced by hydrodynamic tractions on the surface of contacting particles.
These tractions induce a long-ranged velocity field in the fluid, which in turn slightly moves all the other particles in the fluid.
Because contact forces are not time-reversible (as contacts can be pushed on, but not pulled on), the motion induced by hydrodynamic interactions created by contacts is in turn not reversible.
As hydrodynamic interactions are long-ranged, these mediated displacements have been suspected to play a key role in the nature of the RIT~\cite{weijsEmergentHyperuniformityPeriodically2015,mariAbsorbingPhaseTransitions2022}. 

In the stroboscopic point of view of the ROM, the presence of mediated interactions means that passive particles should not be considered motionless, but rather that they are subject to irreversible displacements induced by active particles.
Introducing within a stroboscopic model a detailed description of the motion of passive particle from hydrodynamic interactions that act over a full shear cycle, while in principle possible, somewhat defeats the purpose of having a minimal model. 
It is probably not needed if the goal is to study the critical properties, which we expect will not depend on microscopic interaction details.
Hydrodynamic boundary conditions may be relevant, but studying systematically their effect would be daunting as the first take on the role of hydrodynamic interactions that this work aim at being.

We thus make some modelling choices that we think make a reasonable tradeoff between keeping the relevant ingredient of hydrodynamic interactions during a shear cycle and establishing a minimal model in the ROM spirit which is readily implementable numerically but also lends itself to analytical arguments.
First, we decide to ignore the fact that contacts evolve during the cycle and therefore induce time varying hydrodynamic interactions.
Then, because we expect that the crucial aspect of hydrodynamic interactions is that they are long ranged, we ignore short-ranged features of hydrodynamic interactions and consider only the large distance leading order.
In other words, we assume contacts during a cycle to act as force dipoles, and the induced motion on passive particles is thus governed by the large distance behavior of the Stokesian hydrodynamic mobility functions~\cite{kimMicrohydrodynamicsPrinciplesSelected1991}.

In the original ROM, two active particles in overlap receive random kicks with uncorrelated directions and amplitudes (although a center-of-mass conserving ROM was also studied~\cite{hexnerNoiseDiffusionHyperuniformity2017,wilkenRandomClosePacking2021}). 
To remain consistent with this choice, we consider that the displacements induced on passive particles are also uncorrelated.
We therefore assume that a contact induces a displacement on a passive particle with a direction picked at random and an amplitude that depends on the distance between the contact and the passive particle.
In the same spirit, we consider that the random displacements induced by two different contacts are uncorrelated, so that the variance of the resulting random displacement is additive.

Finally, we make a last choice primarily motivated by the efficiency of the numerical implementation (which can become a limiting factor with long range interactions). 
Instead of considering the effect of all active particles on all passive particles, we use a coarse grained interaction, which is compatible with our intent of keeping only the long-range nature of interactions.
As sketched in \autoref{fig:rules}, we divide our system in a regular cubic grid of boxes of size $\ell$, and define an activity field $A$ such that for box $\mathfrak{b}$ centered on position $\bm{r}_\alpha$ 
\begin{equation}
    A^t_\mathfrak{b} = \frac{1}{l^d}\sum_{i\in \mathfrak{b}} n_i\, .
\end{equation}
where $n_i=0,1$ if particle $i$ is passive or active, respectively.
For each passive particle $i$, we compute the amplitude  $\Delta_{\mathrm{p},i}$ defined through 
\begin{equation}
    \Delta_{\mathrm{p},i} = \sqrt{\sum_{\mathfrak{b}'} G(r_{\mathfrak{b}'\mathfrak{b}}) A_{\mathfrak{b}'}}
    \label{eq:passive_step_size}
\end{equation}
where $\mathfrak{b}$ is the box particle $i$ belongs to, and $r_{\mathfrak{b}'\mathfrak{b}}$ is the distance between the center of boxes $\mathfrak{b}'$ and $\mathfrak{b}$. At each time step, passive particles are subject to a random kick $\bm{\delta}_{p,i}$, each component of which is a Gaussian random variable of zero mean and standard deviation $\Delta_{\mathrm{p},i}$.
In physical terms, the kernel $G(r)$ plays the role of the square of the physical propagator $\mathcal{G}(\mathbf{r})$ encoding the integrated effect of hydrodynamic interactions over a cycle in the cyclically sheared suspension experiment. 
The physical propagator $\mathcal{G}(\mathbf{r})$ 
decays at large distances like the hydrodynamic mobility functions, as $1/r^{\alpha}$, with an exponent $\alpha$ which depends on the boundary conditions and therefore on the experimental setup considered, as already mentioned~\cite{diamantHydrodynamicInteractionConfined2009}.
Thus the kernel $G(r)$ decays at large separations as $1/r^{2\alpha}$.
In practice, we pick 
\begin{equation}
    G(r) = \frac{c}{(1+r^2)^{\alpha}}
    \label{eq:kernel}
\end{equation}
because it has an explicit discrete Fourier transform in $d=2$ and $d=3$, which is convenient to perform the discrete convolution in \autoref{eq:passive_step_size}. 
The prefactor $c$ depends on the interaction range. 
The form prescribed by \autoref{eq:kernel} allows a convergence of the convolution in the limit $L\to\infty$ with a constant $c$ as long as $\alpha > d/2$. 
In this case, we take $c=0.25$, independently of $\alpha$.
For $\alpha < d/2$ however, this convolution diverges as $L^{d-2\alpha}$. 
In order to keep a meaningful $L\to \infty$ limit for these values of $\alpha$, we pick $c = 0.25L^{2\alpha-d}$.

We may expect two limiting cases for the mediated ROM behavior. 
For large enough values of $\alpha>\alpha_\mathrm{SR}$, we should observe a short-range, $\alpha$-independent behavior. 
We will see shortly that in this short-range regime the mediated ROM shares the same critical behavior as the ROM, and therefore the large-$\alpha$ mediated ROM is a CDP-class model.
On the other end of the spectrum, for small $\alpha$ values, we may expect a mean-field-like behavior from the very-long-ranged mediated interactions, similar to what is usually observed for the critical behavior of systems with long-range interactions~\cite{fisherCriticalExponentsLongRange1972,hinrichsenNonequilibriumPhaseTransitions2007}.
In a previous article, we introduced and simulated a modified ROM with mediated interactions treated at a mean-field level, that is, with $\Delta_{\mathrm{p},i} \propto \sqrt{A}$ with $A=N^{-1}\sum_i n_i$ the fraction of active particles, which corresponds to the $\alpha=0$ limit of the mediated ROM. 
We therefore expect the small-$\alpha$ limit of the mediated ROM to behave like this previously studied model.

\section{Numerical results}
\label{sec:numerical:results}

\begin{figure}[h]
	\includegraphics[width=\linewidth]{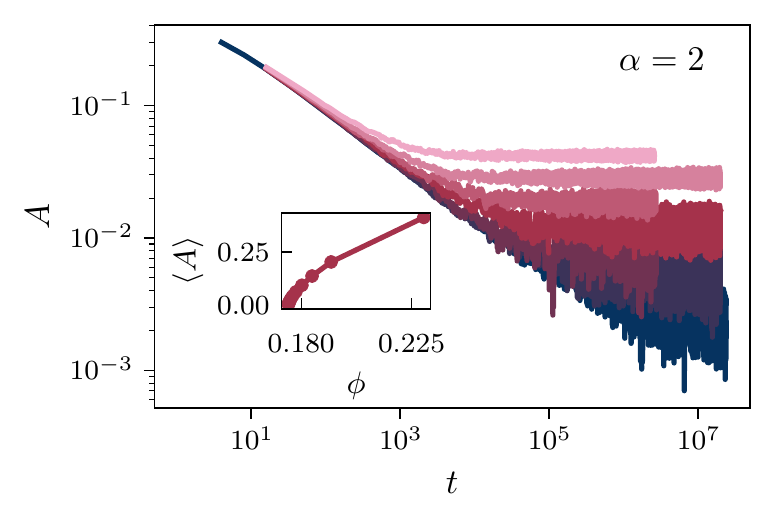}
	\caption{Evolution of the global activity $A(t)$ in the system as a function of time $t$, for a range $\alpha=2$ in dimension $d=2$. Starting from a random initial state, activity decreases to a stationary value.}
	\label{fig:Decay}
\end{figure}

We show the time evolution of $A$ in \autoref{fig:Decay} for $d=2$ and $\alpha=2$ for several values of $\phi$, starting from configurations with particle postitions picked in a random uniform distribution within the box.
Just like the original ROM the mediated ROM exhibits a second order APT at a critical density $\phi_\mathrm{c}$ (which value depends on both $d$ and $\alpha$). 
In this section, we characterize the critical behavior of the mediated ROM primarily in dimension $d=2$ and partially in $d=3$, and as a function of the interaction range $\alpha$. For $d=2$, we consider system sizes up to $L=8192$ and numbers of particles up to $N\approx \num{1e7}$. For $d=3$, we consider system sizes up to $L=768$ and numbers of particles up to $N \approx \num{4e7}$.

\subsection{From concave to convex}

\begin{figure}[h]
	\includegraphics[width=\linewidth]{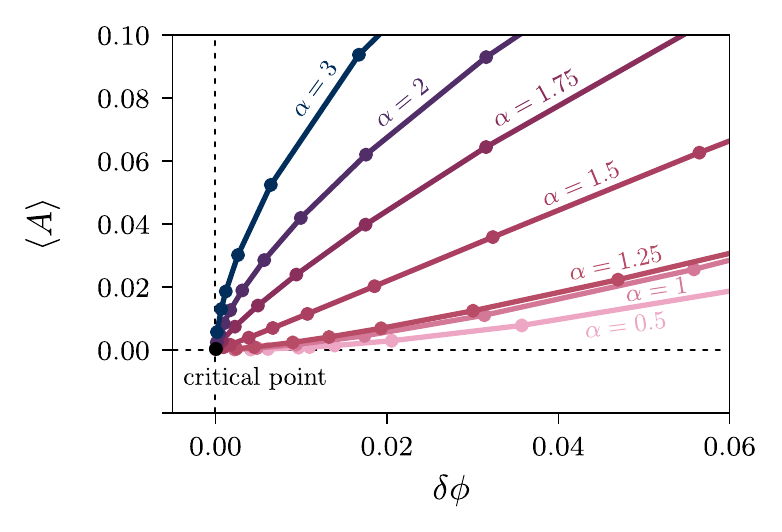}
	\caption{Stationary average activity $\langle A \rangle$ as a function of $\delta\phi = \phi-\phi_\mathrm{c}$ for different ranges of interaction $\alpha$.
	Data are plotted on linear scales to emphasize the change from concave to convex curves when $\alpha$ is decreased.}
	\label{fig:convexity}
\end{figure}

In the stationary state, it is expected that close to the critical density $\phi_\mathrm{c}$ the average activity in the system $\langle A \rangle$ vanishes as
\begin{equation}
	\langle A \rangle \sim \delta\phi^\beta,\quad \delta\phi = \frac{\phi-\phi_\mathrm{c}}{\phi_\mathrm{c}}\, .
	\label{eq:critical_op}
\end{equation}

In \autoref{fig:convexity}, we report the $\phi$ dependence of the steady-state activity for several values of $\alpha$ between $\alpha=0.5$ and $\alpha=3$, for $d=2$.
For the sake of comparison between different values of $\alpha$, we show $\langle A\rangle$ as a function of $\delta\phi$, which of course implies a prior determination of $\phi_\mathrm{c}$ which we will describe shortly.
Qualitatively, we can see that the curve $\langle A \rangle = f(\delta\phi)$ goes from concave for short ranges of interaction (large $\alpha$) to convex for long ranges of interaction (small $\alpha$), the transition point being around $\alpha\approx 1.5$.

\begin{figure}[h]
    \centering
    \includegraphics[width=\linewidth]{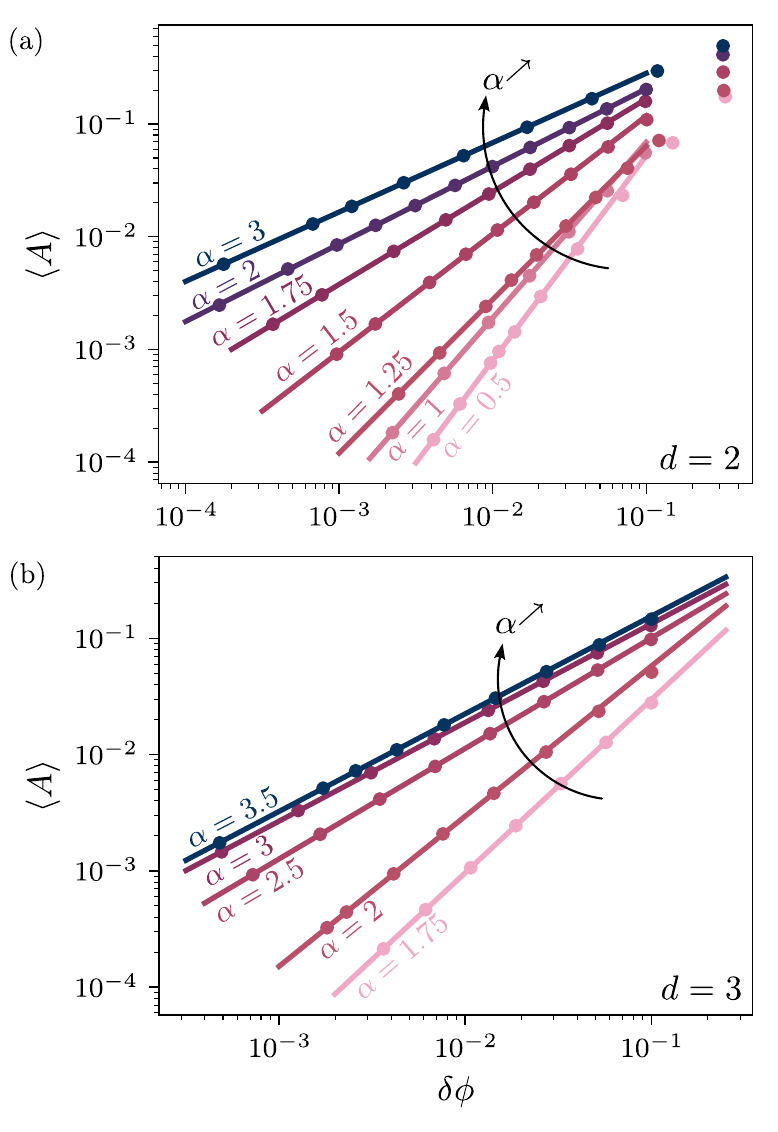} 
    \caption{Log-log plots of the average activity $\langle A \rangle$ in stationary state as a function of $\delta\phi = \phi-\phi_\mathrm{c}$ for different ranges of interaction $\alpha$,
    for d=2 (a) and d=3 (b). Data in panel (a) are the same as in \autoref{fig:convexity}.}
    \label{fig:evolmean}
\end{figure}

To determine more precisely the scaling of $\langle A \rangle$ with $\delta\phi$, we must first accurately determine the critical density $\phi_\mathrm{c}$ associated with each range of interaction $\alpha$. In practice, this is done by plotting $\langle A \rangle = f(\delta\phi)$ and identifying the critical density $\phi_\mathrm{c}$ for which the curve is a straight line at small $\delta\phi$ on a logarithmic scale. 
The $\beta$ exponent is then the slope of this curve. 
The result of this procedure is shown in \autoref{fig:evolmean}(a) for $d=2$ and \autoref{fig:evolmean}(b) for $d=3$. 
The uncertainty on $\phi_\mathrm{c}$ is determined as the range of trial critical densities over which we can see a reasonable straight line, and we define the uncertainty on $\beta$ as the range of slopes measured for this range of critical densities.

For $d=2$, we find for the shortest range $\alpha=3$ an exponent $\beta= 0.62(2)$.
This value is close to the one reported for the ROM~\cite{mariAbsorbingPhaseTransitions2022} and the expected value of $\beta=0.639(9)$ for the CDP class~\cite{lubeckUniversalScalingBehavior2004}.
Similarly, for $d=3$ we find $\beta = 0.84(2)$ for the shortest investigated range $\alpha=3.5$, which is close to the reported $\beta=0.840(12)$ for CDP~\cite{lubeckUniversalScalingBehavior2004}.
These are not coincidences: having active particles affect passive particles only in their immediate neighborhood is not a relevant change to the case where active particles do not affect passive ones at all.

We then find a range of $\alpha$ for which we find floating, $\alpha$-dependent values for $\beta$, as shown in \autoref{fig:exprecap_2d}(a) for $d=2$ and \autoref{fig:exprecap_3d}(a) for $d=3$.
Variations of $\beta$ are especially strong when $\alpha$ is decreased below $\alpha\approx 2$ for $d=2$ and $\alpha\approx 3$ for $d=3$.
While this may suggest that $\alpha=d$ is the upper bound for floating exponents above which the model  is short-ranged and falls into CDP, the exact value of this bound and its dependence on $d$ cannot be firmly established based on the coarse sampling in $\alpha$ we investigated here.
Further decreasing $\alpha$ values, the transition turns convex for $\alpha\approx 1.5$ in $d=2$, and  $\alpha\approx 2.5$ in $d=3$.
In $d=2$, we could measure an increase of $\beta$ up to $\beta=1.82(15)$ for $\alpha=0.5$.
This value is close to the $\beta\approx 1.85$ reported for the case of mean-field-level mediated interactions~\cite{mariAbsorbingPhaseTransitions2022}, which suggests that for $\alpha\lesssim 0.5$ the behavior of the transition is $\alpha$-independent  when $d=2$.

\subsection{From diverging to vanishing fluctuations}

We now turn to the critical behavior of the fluctuations of the order parameter $\langle \delta A^2\rangle = N \langle [A - \langle A\rangle]^2\rangle$, which are expected to scale as
\begin{equation}
	\langle \delta A^2 \rangle \sim \delta\phi^{-\gamma^\prime}\, .
\label{eq:critical_fluctuations}
\end{equation}

For the CDP class, $\gamma^\prime > 0$, leading to diverging fluctuations of the activity as the critical point is approached. 
By contrast, RIT with mean-field-like mediated interactions were argued to exhibit $\gamma^\prime<0$, that is, vanishing critical fluctuations~\cite{mariAbsorbingPhaseTransitions2022}.

\begin{figure}[h]
    \centering
    \includegraphics[width=\linewidth]{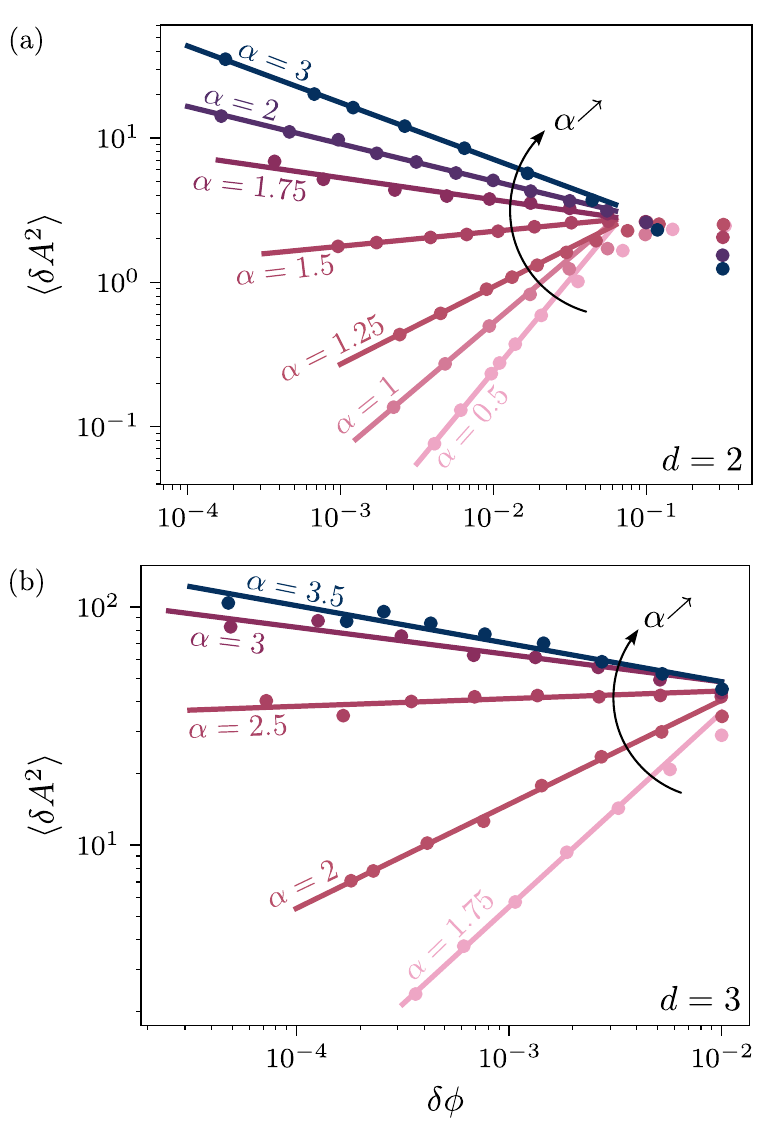} 
    \caption{Variance of activity fluctuations $\langle \delta A^2 \rangle$ in stationary state as a function of $\delta\phi = \phi-\phi_\mathrm{c}$ for different ranges of interaction $\alpha$, for $d=2$ (a) and $d=3$ (b).}
    \label{fig:evolmeanvar}
\end{figure}

The variance $\langle \delta A^2\rangle$ of activity fluctuations is plotted in \autoref{fig:evolmeanvar}(a) for $d=2$ and \autoref{fig:evolmeanvar}(b) for $d=3$, for several values of $\alpha$,
showing a transition from diverging to vanishing fluctuations when $\alpha$ is decreased.
We report the $\alpha$ dependence in \autoref{fig:exprecap_2d}(b) for $d=2$ and \autoref{fig:exprecap_3d}(b) for $d=3$.
We find that the $\gamma^\prime$ exponent is floating (i.e., $\alpha$-dependent) in the same ranges of $\alpha$ values as the $\beta$ exponent.
Starting from the short-range limit (large $\alpha$), in $d=2$ for $\alpha\gtrsim 2$, we find $\gamma^\prime = 0.39(3)$, a value compatible with the value $\gamma^\prime = 0.367(19)$ reported for CDP~\cite{lubeckUniversalScalingBehavior2004}.
In $d=3$, for $\alpha\gtrsim 3$ we find $\gamma^\prime=0.16(6)$, here again in agreement with the CDP value $\beta= 0.152(17)$~\cite{lubeckUniversalScalingBehavior2004}.
Decreasing $\alpha\lesssim d$, we observe a decrease of $\gamma^\prime$, with $\gamma^\prime$ changing sign from positive to negative for $\alpha\approx 1.5$ in $d=2$, and  $\alpha\approx 2.5$ in $d=3$, that is, concurrently with the transition turning from concave to convex.
(Intriguingly, this concurrent change of behavior has also been observed in the case of the yielding transition in elasto-plastic models~\cite{jocteurYieldingAbsorbingPhase2024a}.)
Further decreasing the value of $\alpha$, for $\alpha =0.5$ in $d=2$, we measure a value $\gamma^\prime \approx -1.28(15)$ compatible with the one reported for mean-field-like mediated interactions~\cite{mariAbsorbingPhaseTransitions2022}.

Usually, the evolution of the mean value of the order parameter and of its fluctuations are not independent. Indeed, in most phase transitions there exists a scaling relation between the exponent $\beta$, $\gamma^\prime$ and $\nu_\perp$ the exponent characterizing the divergence of the correlation length $\xi \sim \delta\phi^{-\nu_\perp}$,
\begin{equation}
	2\beta + \gamma^\prime = \nu_\perp d\, ,
	\label{eq:scaling_relation}
\end{equation}
called hyperscaling relation.
Considering the fact that this relation was shown to hold in the short-range limit $\alpha\rightarrow \infty$ of CDP \cite{lubeckUniversalScalingBehavior2004} and in the long-range limit $\alpha = 0$ \cite{mariAbsorbingPhaseTransitions2022}, one might expect that it holds for any $\alpha$ in between. 
Under this assumption, it is possible to derive a correlation length exponent $\nu_\perp^*$ for every $\alpha$ based on the previous measurements of $\beta$ and $\gamma^\prime$. 
(Here we use the $\nu_\perp^*$ notation instead of $\nu_\perp$ to highlight the fact that this is not directly a measured value, but merely a by-product of two other measured exponents assuming hyperscaling.)
The computed values, reported in \autoref{table:exponents} for $d=2$ and $d=3$, show much smaller relative variations of $\nu_\perp^\ast$, of order \SI{50}{\percent} over the range of $\alpha$ investigated for $d=2$, compared to the relative variations of $\beta$ and $\gamma^\prime$ on the same range.

\subsection{Relaxation towards the stationary state}

\begin{figure}[h]
    \centering
    \includegraphics[width=0.48\textwidth]{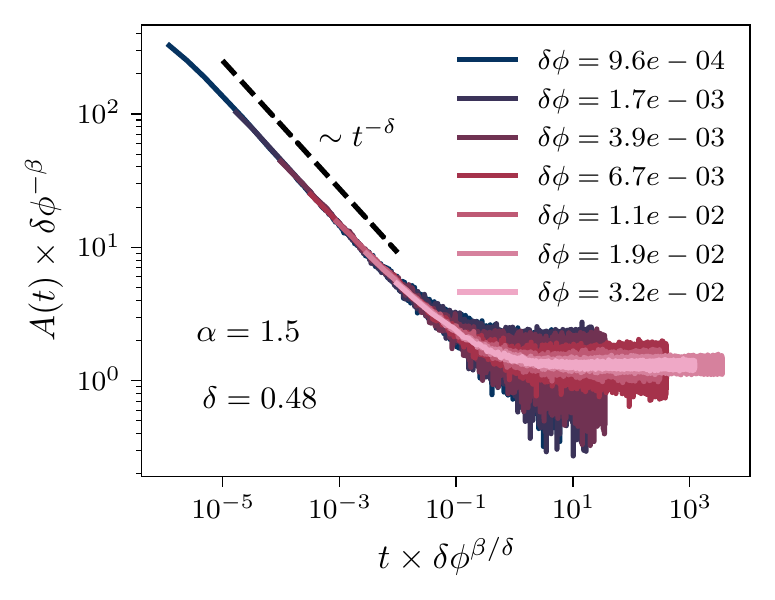}
    \caption{Rescaled time evolution of the activity $A(t)$ for different $\delta\phi = \phi-\phi_\mathrm{c}$, for $\alpha=1.5$ and $d=2$. The collapse of the curves determines $\delta = 0.49(1)$.}
    \label{fig:deltadeter}
\end{figure}

We now turn to the activity relaxation exponent. 
Right at $\phi=\phi_\mathrm{c}$, we expect the activity to decay algebraically in time as
\begin{equation}
	A(t) \sim t^{-\delta}\, .
	\label{eq:critical_decay}
\end{equation}

We measure this dynamical exponent through a rescaling of the relaxation curves for $A$. 
Indeed, if the scalings in \autoref{eq:critical_op} and \autoref{eq:critical_decay} hold, we expect that the $A(t)$ curves for several values of $\delta\phi$ should collapse onto a single master curve under the rescaling $A(t)\delta\phi^{-\beta} = h(t\delta\phi^{\beta/\delta})$.
The transition point $\phi_\mathrm{c}$ and the exponent $\beta$ being determined previously, 
$\delta$ can be determined as the exponent which achieves the collapse. 
An example of determination of $\delta$ for $\alpha=1.5$ and $d=2$ is given in \autoref{fig:deltadeter}. (Here we performed the collapse for $\phi>\phi_\mathrm{c}$, but a similar collapse is expected to hold with the same exponents for $\phi<\phi_\mathrm{c}$, only on a different master curve.) 
The error on the determined exponent $\delta$ is defined as the range of values for which we can achieve a collapse.

\begin{figure}[h]
    \centering
    \includegraphics[width=\linewidth]{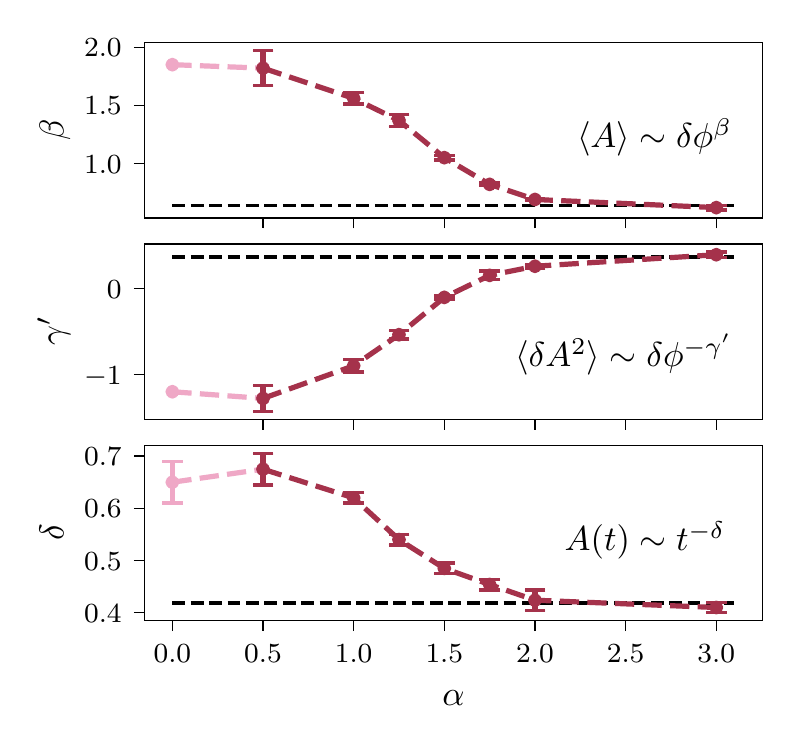}
    \caption{Critical exponents $\beta$, $\gamma^\prime$ and $\delta$ (see text for definitions) in the mediated ROM as a function of the mediated interaction range exponent $\alpha$, for dimension $d=2$. Pale pink dots represent the results obtained in \cite{mariAbsorbingPhaseTransitions2022} for $\alpha=0$.}
    \label{fig:exprecap_2d}
\end{figure}

\begin{figure}[h]
    \centering
    \includegraphics[width=\linewidth]{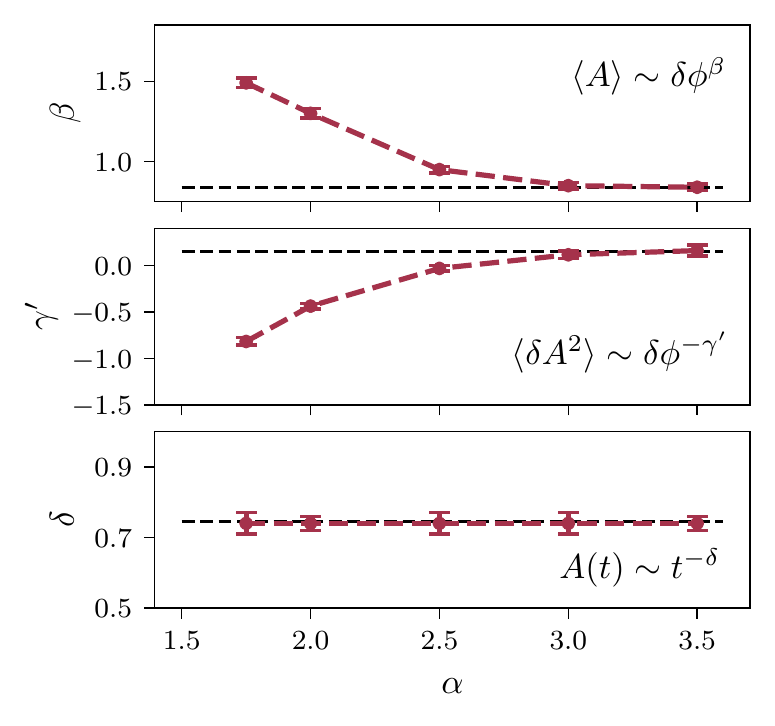}
    \caption{Critical exponents $\beta$, $\gamma^\prime$ and $\delta$ (see text for definitions) in the mediated ROM as a function of the mediated interaction range exponent $\alpha$, for dimension $d=3$.}
    \label{fig:exprecap_3d}
\end{figure}

With this method we are able to determine $\delta$ for all the values of $\alpha$ we investigated. We show the results in \autoref{fig:exprecap_2d}(c) for $d=2$ and \autoref{fig:exprecap_3d}(c) for $d=3$. 
We once again find the same range of $\alpha$ over which $\delta$ varies continuously.
For $\alpha \gtrsim d$, we measure $\delta \approx 0.42$ for $d=2$ and $\delta\approx 0.74$ for $d=3$.
These values are in good agreement with the reported values for CDP, which are $\delta=0.419 \pm 0.015$ for $d=2$ and $\delta=0.745\pm 0.017$ for $d=3$~\cite{lubeckUniversalScalingBehavior2004}. 
In  $d=2$, decreasing the value of $\alpha$ below $\alpha=2$, we find that the value of $\delta$ increases, up to a $\delta=0.67$(3) for $\alpha=0.5$.
Surprisingly we do not see a similar behavior for $d=3$, for which we do not observe any significant change in the value of $\delta$ down to $\alpha=1.75$.

\begin{table*}[t]
\caption{Measured critical exponents for the mediated ROM in $d=2$ (top rows) and $d=3$ (bottom rows), except for $\nu_\perp^\ast$, which value is deduced via hyperscaling, \autoref{eq:scaling_relation}.}
\begin{ruledtabular}
\begin{tabular}{>{\centering\arraybackslash}p{0.04\textwidth} >{\centering\arraybackslash}p{0.16\textwidth} >{\centering\arraybackslash}p{0.16\textwidth} >{\centering\arraybackslash}p{0.16\textwidth} >{\centering\arraybackslash}p{0.16\textwidth} >{\centering\arraybackslash}p{0.16\textwidth} >{\centering\arraybackslash}p{0.16\textwidth}} 
 & $\alpha$ & \multicolumn{1}{c}{$\beta$} &
\multicolumn{1}{c}{$\gamma^\prime$}&
\multicolumn{1}{c}{$\delta$} &
\multicolumn{1}{c}{$\nu_\perp^\ast$} &
\multicolumn{1}{c}{$\theta$} \\
 \hline
  \multirow{9}{*}{\rotatebox[origin=c]{90}{$d=2$}} & CDP~\cite{lubeckUniversalScalingBehavior2004} & 0.639(9) & 0.367(19) & 0.419(15)$^a$ & 0.799(14) & - \\
 & 3 & 0.62(2) & 0.39(3) & 0.41(1) & 0.815(35) & 0.57(3) \\ 
 & 2 & 0.69(1) & 0.26(2) & 0.42(2) & 0.82(2) & 0.78(5) \\
 & 1.75 & 0.82(1) & 0.15(5) & 0.45(2) & 0.895(35) & 0.76(5) \\
& 1.5 & 1.05(2) & -0.10(2) & 0.49(1) & 1.00(3) & 0.79(10) \\
& 1.25 & 1.37(5) & -0.54(5) & 0.54(1) & 1.10(7) & 0.81(5) \\
& 1 & 1.56(3) & -0.90(7) & 0.62(1) & 1.11(7) & 0.86(3) \\
& 0.5 & 1.82(15) & -1.28(15) & 0.67(3) & 1.18(22) & 0.91(5) \\
& 0 \cite{mariAbsorbingPhaseTransitions2022} & 1.85 & -1.2 & 0.65(4) & 1.3 & - \\
 \hline
 \multirow{6}{*}{\rotatebox[origin=c]{90}{$d=3$}} & 
 CDP~\cite{lubeckUniversalScalingBehavior2004} & 0.840(12) & 0.152(17) & 0.745(17)$^a$ & 0.593(13) & - \\
 & 3.5 & 0.84(2) & 0.16(6) & 0.74(2) & 0.61(3) & - \\ 
 & 3 & 0.85(2) & 0.11(4) & 0.74(3) & 0.60(3) & - \\
 & 2.5 & 0.95(2) & -0.03(3) & 0.74(3) & 0.62(2) & - \\
 & 2 & 1.30(3) & -0.44(3) & 0.74(2) & 0.72(3) & - \\
 & 1.75 & 1.49(3) & -0.82(4) & 0.74(3) & 0.72(3) & -
 \label{table:exponents}
\end{tabular}
\end{ruledtabular}
\begin{flushleft}
\footnotesize{$^a$ The exponent $\delta$ is called $\alpha$ in~\cite{lubeckUniversalScalingBehavior2004}.}
\end{flushleft}
\end{table*}

\subsection{From hyperuniform to uniform}
\label{sec:hyperunif}

\begin{figure}[h]
    \centering
    \includegraphics[width=\linewidth]{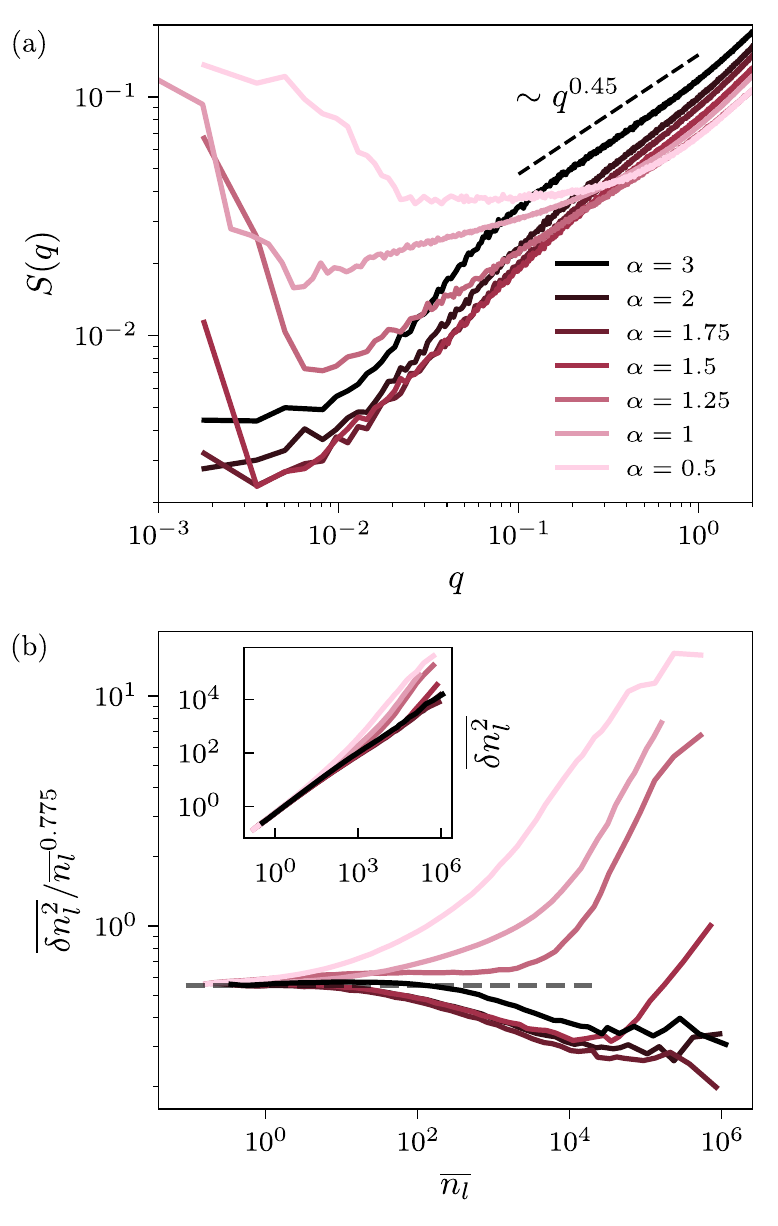}
    \caption{Density fluctuations for $d=2$ and for several values of $\alpha$, at a distance $\delta \phi\approx \num{e-2}$ from the critical point. (a) Structure factor $S(q)$. (b) Rescaled number fluctuations $\overline{\delta n_l^2}/\overline{n_l}^{0.775}$ as a function of $\overline{n_l}$, obtained from the box counting algorithm.}
    \label{fig:HUdphi4e3}
\end{figure}

A remarkable CDP-class feature, which attracted a lot of attention lately, is hyperuniformity at the absorbing phase transition.
We introduce the structure factor
\begin{equation}
S(\bm{q}) = N^{-1} \left\langle \sum_{i, j} e^{\mathrm{i} \bm{q}\cdot(\bm{r}_i - \bm{r}_j)} \right\rangle\, .
\end{equation}
As the structure factor is isotropic for the models we consider in this article, in the following we denote it as $S(q)$, with $q=|\bm{q}|$.
Looking at the small-$|\bm{q}|$ behavior, we can define an exponent $\zeta$ as
\begin{equation}
S(q) \underset{q\to 0}{\sim} q^\zeta\, .
\label{eq:sfact_scaling}
\end{equation}
Uniform equilibrium systems have normal large-scale density fluctuations due to thermal fluctuations, with $\zeta=0$.
The ROM model, and more generally models in the CDP class, instead show hyperuniformity at $\phi=\phi_\mathrm{c}$, that is, have suppressed large-scale density fluctuations corresponding to $\zeta>0$, with a reported value $\zeta\approx 0.45$ in $d=2$~\cite{tjhungHyperuniformDensityFluctuations2015,hexnerHyperuniformityCriticalAbsorbing2015,hexnerNoiseDiffusionHyperuniformity2017}.

An alternative way to probe density fluctuations is to perform a box counting analysis~\cite{torquatoLocalDensityFluctuations2003}. 
We consider the number of particles $n_l(\bm{r})$ in a cubic box of size $l$ centered in $\bm{r}$. 
This number depends on $\bm{r}$ and is thus fluctuating around the average value $\overline{n_l} = N(l/L)^d$. 
Introducing its variance $\overline{\delta n_l^2} = \overline{[n_l - \overline{n_l}]^2}$, where $\overline{\bullet}$ denotes an average over $\bm{r}$, a hyperuniform system is characterized by the scaling
\begin{equation}
    \overline{\delta n_l^2} \underset{l\to \infty}{\sim} (\overline{n_l})^{\eta},\quad \eta < 1\, .
\label{eq:nfluct_scaling}
\end{equation}
By contrast, normal fluctuations correspond to $\eta=1$. 
Characterizing density fluctuations via the small-$q$ behavior of the structure factor or the large-$l$ behavior of the number variance is equivalent in the limit of large systems, and we have $\eta = 1-\zeta/d$~\cite{torquatoHyperuniformStatesMatter2018}.
For the ROM in $d=2$, this implies $\eta\approx 0.775$, which is indeed consistent with measured values~\cite{tjhungHyperuniformDensityFluctuations2015,hexnerHyperuniformityCriticalAbsorbing2015}.

In practice, for an APT critical properties at $\phi=\phi_\mathrm{c}$ are challenging to observe numerically because finite-size systems always fall in an absorbing state in a finite time due to fluctuations~\cite{lubeckScalingBehaviorOrder2002}. 
One is then forced to simulate systems slightly away from the transition in the active phase, and to take advantage of the critical scaling to infer the behavior at $\phi=\phi_\mathrm{c}$.
For values of $\phi>\phi_\mathrm{c}$ close but not exactly at the critical point, hyperuniformity does not extend to the largest length scales $l$. 
Instead it is observed only in an intermediate range of $l$, below some cutoff length sometimes called hyperuniformity length scale $\ell_\mathrm{HU}$~\cite{tjhungCriticalityCorrelatedDynamics2016} which is also a critical quantity diverging at $\phi_\mathrm{c}$.
This means that the scaling in \autoref{eq:sfact_scaling} should hold only for $2\pi/\ell_\mathrm{HU}^d \ll q \ll 1$ and the one in \autoref{eq:nfluct_scaling} should hold only for $1 \ll \overline{n_l} \ll N(\ell_\mathrm{HU}/L)^d$.

To enable comparison of the density fluctuations between systems with different ranges of interaction $\alpha$, 
we then perform our analysis with roughly the same given value of $\delta \phi\approx \num{e-2}$ for all $\alpha$. 
In \autoref{fig:HUdphi4e3}(a) we show the $d=2$ results for the structure factor $S(q)$.
For $\alpha=3$, we observe a behavior similar to the one of the usual ROM.
As we are at a finite distance from the critical point, we do observe the $q^{0.45}$ scaling typical of CDP hyperuniformity in an intermediate range of $q=\numrange{e-1}{1}$.
For $q<\num{e-1}$, the slope is larger than $0.45$, indicating a system with even lower density fluctuations than at the critical point. 
This phenomenon is well-known in the ROM~\cite{hexnerHyperuniformityCriticalAbsorbing2015}.
These suppressed density fluctuations do not extend to arbitrarily large length scales however, as for $q<\num{e-2}$ the structure factor plateaus, indicating that normal fluctuations ($\zeta=0$) are recovered for the largest length scales.
The location of the crossover between this plateau and the range of suppressed density fluctuations is related to $\ell_\mathrm{HU}$, and we expect that it collapses at $\phi_\mathrm{c}$, leaving only the hyperuniform regime~\cite{hexnerHyperuniformityCriticalAbsorbing2015,hexnerNoiseDiffusionHyperuniformity2017}.

Now focusing on values of $\alpha<3$, that is, on the range of $\alpha$ for which the critical exponents studied so far have floating values, we observe only minute differences for $S(q)$ down to $\alpha=1.5$, in fact corresponding to a small global decrease of $S(q)$ irrespective of the value of $q$.
The first marked change occurs at $\alpha=1.5$, for which we observe that the value of $S(q)$ at the smallest $q$ in our system is almost an order of magnitude above the value at the next smallest $q$.
Looking at the data for even smaller $\alpha$ values, we attribute this steep variation as coming from the rise of a high plateau at small $q$.
Concurrent to this plateau extending at increasing values of $q$ when $\alpha$ decreases, we observe a gradual disappearance of the hyperuniform region scaling as $q^{0.45}$.
Crucially, this plateau is not a finite-$\delta\phi$ effect nor a finite-$L$ effect, as shown in \autoref{fig:Sq_Appendix}(a) and \autoref{fig:Sq_Appendix}(b)
in Appendix~\ref{sec:appendix:sfact}, where we observe that $S(q)$ is insensitive to roughly an order of magnitude change in the values of $\delta\phi$ and $L$.
Hyperuniformity is therefore lost, even at the absorbing phase transition, for $\alpha \lesssim 1.5$, which interestingly is also the point at which the transition changes convexity.

The same phenomenon can be observed from the number fluctuations.
In \autoref{fig:HUdphi4e3}(b) we show the $d=2$ results for $\overline{\delta n_l^2}/\overline{n_l}^{0.775}$ as a function of $\overline{n_l}$ for the same systems as shown in \autoref{fig:HUdphi4e3}(a) (in inset we also plot the unrescaled $\overline{\delta n_l^2}$ versus $\overline{n_l}$).
Within this representation, models in the CDP class should fall on a horizontal straight line for an intermediate range of $1 \ll \overline{n_l} \ll N(\ell_\mathrm{HU}/L)^d$.
A model which shows hyperuniformity but with a different value for the exponent $\eta$ would fall on a straight but tilted line.
We observe a horizontal plateau for $\alpha=3$ between $\overline{n_l}\approx 1$ and $\overline{n_l}\approx \numrange{e2}{e3}$ (corresponding to the intermediate range of $q$ showing hyperuniformity in \autoref{fig:HUdphi4e3}(a)), followed by over-suppressed fluctuations for $\overline{n_l} \gtrsim \num{e3}$. 
We do not observe a large-$\overline{n_l}$ uptake for the range of $\overline{n_l}$ that corresponds to the small-$q$ plateau in $S(q)$ in \autoref{fig:HUdphi4e3}(a), rather merely a plateau.
Decreasing the value of $\alpha$, density fluctuations slightly decrease for $\alpha=2$ and $\alpha=1.75$ compared to $\alpha=3$. 
The first significant change we observe is for $\alpha=1.5$, where the density fluctuations now steeply increase at the largest values of $\overline{n_l}$, which is the twin of the small-$q$ bump observed in $S(q)$, and the signature of the loss of hyperuniformity.
For even smaller values of $\alpha$, the large $\overline{n_l}$ rise extends to lower $\overline{n_l}$ values, such that we do not see any range of $\overline{n_l}$ for which data fall on a straight line, let alone on a horizontal one.

When increasing the range of interactions (i.e., decreasing $\alpha$), we thus successively observe two distinct behaviors. When $\alpha$ decreases, we first observe a slight decrease of the amplitude of large-scale fluctuations, before this amplitude strongly increases, starting from the largest length scales and progressively covering a broader range of scales,
see \autoref{fig:HUdphi4e3}(a) and \autoref{fig:HUdphi4e3}(b).
Although we do not have a theoretical description of this phenomenon at hand, these observations suggest the following heuristic argument.
Evaluating the structure factor for a fluid of Brownian particles from fluctuating hydrodynamic equations, one finds $S(q)\propto \sigma^2/\mathcal{D}$,
where $\sigma^2$ is the amplitude of the noise on particle current, and $\mathcal{D}$ is the particle diffusion coefficient. Unsurprisingly, the structure factor increases with increasing noise,
but it also decreases when increasing the diffusion coefficient. The mediated diffusion of passive particles in our model qualitatively combines these two effects, as it both enhances particle diffusion, and is also expected to increase the noise on particle current, potentially leading to competing effects. An expected difference with Brownian particles is that long-range interactions 
may induce long-range correlation in the hydrodynamic noise, which in turn impacts the structure factor with an excess of fluctuations in large-scale modes.

The fact that hyperuniformity disappears when increasing the interaction range in the mediated ROM is consistent with observations from experiments and direct numerical simulations of dilute emulsions confined in a quasi-twodimensional microfluidic channel and subject to oscillatory pressure-driven flow~\cite{weijsEmergentHyperuniformityPeriodically2015}. 
However, in the mediated ROM the range of $\alpha$ at which hyperuniformity diappears is $\alpha\lesssim 1.5$, whereas in the experiments of~\citet{weijsEmergentHyperuniformityPeriodically2015}, there is no visible hyperuniformity on a confined monolayer ($\alpha=2$).
Conversely, our results also stand in contrast to the experimental observations of~\citet{wilkenHyperuniformStructuresFormed2020}, who report that configurations at the RIT of bulk suspensions ($\alpha=d-1$) are hyperuniform (which is also in contradiction with the results of~\cite{weijsEmergentHyperuniformityPeriodically2015}). 
We conjecture that a possible origin for this discrepancy is that the loss of hyperuniformity due to hydrodynamic interactions occurs at a length scale larger than what was investigated in~\cite{wilkenHyperuniformStructuresFormed2020}.
Indeed, in \autoref{fig:HUdphi4e3}(a), for $\alpha=d-1=1$, we see a minimum in $S(q)$ slightly below $qD = \num{e-2}$, whereas the smallest accessible $q$ in~\cite{wilkenHyperuniformStructuresFormed2020} is $qD\approx \num{e-1}$.

\subsection{Where do actual systems stand?}

Our results show that long-range hydrodynamics has a drastic effect on the critical properties.
Crucially, the dependence of the critical exponents on the interaction range exponent $\alpha$ is strong for values of $\alpha$ below $\alpha\approx d$.

Let us recall that the hydrodynamic mobility functions associated with a force dipole source such as a contact during a shear cycle are expected to decay at large distances $r$ as $1/r^{d-1}$ in a bulk system~\cite{kimMicrohydrodynamicsPrinciplesSelected1991}.
For a three-dimensional system, such as the one considered by~\citet{pineChaosThresholdIrreversibility2005}, the relevant value is thus $\alpha=2$, for which we predict a non-CDP, convex transition without hyperuniformity.
These results challenge the interpretation that has commonly be given to these experiments.

By contrast, we predict that for confined systems for which the hydrodynamic mobility decays as $1/r^d$~\cite{diamantHydrodynamicInteractionConfined2009}, such as the one considered by~\citet{weijsEmergentHyperuniformityPeriodically2015}, the transition is within CDP, and therefore concave, with hyperuniformity at the transition.


\section{Theoretical interpretations}

The results presented in the previous section highlight the crucial role of long-ranged hydrodynamic interactions in the critical behavior near the reversible-irreversible transition of cyclically sheared suspensions. 
We first argue that this behavior cannot be understood from a continuum description approach based on extensions of the usual CDP field equations to include long-range transport, before turning to a mean-field approach of the activity creation by diffusion induced at long range.

\subsection{Continuum description}

\subsubsection{A natural benchmark: Long-range CDP field equations}

The fact that critical exponents depend continuously on the exponent $\alpha$ characterizing the decay of long-range interactions in some range of $\alpha$ values is quite generic and should not come as a surprise.
However, the way exponent values depend on $\alpha$ in the mediated ROM does not follow the usual scenario inherited from equilibrium transition~\cite{fisherCriticalExponentsLongRange1972}, as briefly discussed below.

In equilibrium transitions in the presence of long-range interactions, exponent values depend continuously on $\alpha$ in a range $\alpha_\mathrm{MF} < \alpha <\alpha_\mathrm{SR}$~\cite{fisherCriticalExponentsLongRange1972}.
For $\alpha>\alpha_\mathrm{SR}$ (with $\alpha_\mathrm{SR} = d+2-\eta_\mathrm{SR}$, $\eta_\mathrm{SR}$ being the Fisher exponent), the critical properties are the same as an equivalent system with short-range interactions, whereas for $\alpha<\alpha_\mathrm{MF}$ (with $\alpha_\mathrm{MF}>d$), the transition is mean field.
This scenario carries over to many non-equilibrium transitions~\cite{hinrichsenNonequilibriumPhaseTransitions2007}.
In the case of absorbing phase transitions, it is observed when the activity \emph{transport} is long-ranged, thanks for instance to L\'{e}vy flights of active units~\cite{janssenLevyflightSpreadingEpidemic1999,hinrichsenModelAnomalousDirected1999}.

At the continuum level, long-range transport can be captured via fractional derivative terms in the dynamic equations~\cite{hinrichsenNonequilibriumPhaseTransitions2007}.
In the case of CDP, which is described with two fields, a non-conserved activity field $a(\bm{r},t)$ and a conserved density field $\rho(\bm{r},t)$, this gives
\begin{align}
\partial_t \rho & = D \nabla^2 a - D_\mathrm{LR}|\nabla|^{\alpha^\prime-d} a \label{eq:LR_CDP_rho}\\
\partial_t a    & = (r_0+r_1 \rho) a - s a^2  + D \nabla^2 a - D_\mathrm{LR}|\nabla|^{\alpha^\prime-d} a + \sigma \sqrt{a}\eta\, ,
\label{eq:LR_CDP_a}
\end{align}
where $D$, $D_\mathrm{LR}$, $r_0$, $r_1$ and $s$ are positive constants. 
Usual CDP corresponds to $D_\mathrm{LR}=0$~\cite{menonUniversalityClassReversibleirreversible2009}. 
Note that for CDP the transport coefficients $D$ and $D_\mathrm{LR}$ may differ for the $a$ and $\rho$ dynamics, but in the simplest particle models in the CDP class such as the ROM, transport is only achieved by motion of active particles, so that transport coefficients for $a$ and $\rho$ coincide~\cite{wieseCoherentstatePathIntegral2016}.

The fractional derivative operator $|\nabla|^{\alpha^\prime-d}$ is defined through its action on a plane wave $|\nabla|^{\alpha^\prime-d}e^{\mathrm{i}\bm{k}\cdot\bm{r}}=|\bm{k}|^{\alpha^\prime-d}e^{\mathrm{i}\bm{k}\cdot\bm{r}}$. 
This term naturally arises when active particles are subject to displacements given by a vector $\bm{\xi}$ distributed according to $P(\bm{\xi})$ such that $P(\bm{\xi})\sim |\bm{\xi}|^{-\alpha^\prime}$.
Note that in this case $\alpha^\prime$ intervenes in a transport process a priori 
different from the induced diffusion involving the exponent $\alpha$ that we consider in the rest of this article, hence the different notation we use.
Yet, one may wonder whether activity creation arising from induced diffusion might also be accounted for by a continuum theory analogous to \autoref{eq:LR_CDP_rho} and \autoref{eq:LR_CDP_a}.
Indeed, this activity creation will involve the variance of the passive displacements~\cite{mariAbsorbingPhaseTransitions2022}, which from~\autoref{eq:passive_step_size} is nothing but $\propto -|\nabla|^{2\alpha-d} a$.
Such a term will therefore appear in some form in the activity dynamics \autoref{eq:LR_CDP_a} (suggesting $\alpha^\prime = 2\alpha$) although devising what form is unclear at this stage.
Fractional Laplacians will also appear in the conserved field dynamics, but in a form that differs from \autoref{eq:LR_CDP_rho}, as we discuss in the next section.

Scaling analysis applied to \autoref{eq:LR_CDP_rho} and \autoref{eq:LR_CDP_a}  shows that the fractional Laplacians are relevant terms only for $3d/2<\alpha^\prime<d+2$. 
For $\alpha^\prime>\alpha^\prime_\mathrm{SR} = d+2$ the usual Laplacian $D \nabla^2 a$ is relevant and the transition belongs to CDP.
For $\alpha^\prime < \alpha^\prime_\mathrm{MF} = 3d/2$, the nonlinear term $sa^2$ is relevant and the transition follows mean-field directed percolation (the conserved quantity does not affect the mean-field behavior).
In the range $3d/2<\alpha^\prime<d+2$, values of the critical exponents then continuously evolve from the CDP values at $\alpha^\prime=d+2$ to the mean-field directed percolation values at $\alpha^\prime=3d/2$.
Under the assumption that activity creation from induced diffusion corresponds to $\alpha^\prime = 2\alpha$, the CDP universality would be unstable with respect to mediated interactions for $\alpha<(d+2)/2$, that is, $\alpha=2$ for $d=2$ and $\alpha=5/2$ for $d=3$, which are values compatible with what we observe numerically, here stressing again that our resolution on $\alpha$ is rather coarse.
On the other end of the floating exponent range however, with \autoref{eq:LR_CDP_rho} and \autoref{eq:LR_CDP_a} one should recover mean-field CDP, with $\beta=1$ and $\gamma^\prime=0$, at $\alpha=3d/4$, which is $\alpha=3/2$ for $d=2$ and $\alpha=9/4$ for $d=3$.
Instead, in the mediated ROM we observe a convex transition with vanishing fluctuations for $\alpha\lesssim 1.5$ for $d=2$ and for $\alpha\lesssim 2.5$ for $d=3$, see \autoref{table:exponents}.
This discrepancy 
implies that the behavior of the mediated ROM cannot be interpreted as a straightforward generalized CDP field description with effective long-range transport.
Instead, in the next section, we argue that an extra channel of activity creation must be added to generalized CDP.

\subsubsection{CDP field equations with long-range mediated diffusion}

The long-range CDP equations described above assume an activity creation channel schematically represented as $A+P\to 2A$ with long-range hoppings (i.e., activity transport),
where $A$ and $P$ stand for active and passive particles respectively.
In the mediated ROM, active particles actually do not perform long-range hoppings, so this long-range transport of activity is irrelevant. 
Instead, the long-range nature of the model intervenes in the \emph{induction} of activity, that is $2P\to 2A$ via a long-range mediated noise, and we will argue that long-range induced activity indeed generically lead to a convex transition.
(Long-range transport may also lead to convex transitions, but only when the L\'evy flights are cut off at the nearest active site~\cite{ginelliContactProcessesLong2006,hinrichsenNonequilibriumPhaseTransitions2007,argoloVanishingOrderparameterCritical2013}, which is not what happens in the mediated ROM.) 

Including the diffusion of passive particles induced by long-range interactions with active particles in CDP field equations, we obtain
\begin{equation}
\partial_t \rho  = D \nabla^2 a + \nabla \big( (G*a) \nabla (\rho-a) \big)\, , \label{eq:mediated_CDP_rho}
\end{equation}
where $(\rho-a)$ is the local density of the passive particles, and
\begin{equation}
    G*a(\bm{r},t) = \int d\bm{r}'\, G(\bm{r}-\bm{r}')\, a(\bm{r}',t)
\end{equation}
is the convolution of the long-range propagator $G(\bm{r}) \sim 1/r^{2\alpha}$ with the activity field $a(\bm{r},t)$, which describes induced diffusion.
If we further assume that the activity follows the usual CDP dynamics
\begin{equation}
\partial_t a = (r_0+r_1 \rho) a - s a^2  + D \nabla^2 a + \sigma \sqrt{a}\eta\, ,
\label{eq:mediated_CDP_a}
\end{equation}
we can perform a scaling analysis in a similar way as for the long-range CDP case.
We find that diffusion induced at long range is relevant only if both $\alpha<1$ and $d<2$, otherwise the regular diffusion of active particles dominates.
Thus \autoref{eq:mediated_CDP_rho} and \autoref{eq:mediated_CDP_a} seem unable to account for the observed behavior of the mediated ROM for $\alpha\lesssim d$, although the behavior of these equations is consistent with the numerical observation of short-range CDP exponents for $\alpha \gtrsim d$ in the mediated ROM.

The most probable culprit is the activity dynamics, \autoref{eq:mediated_CDP_a}, which does not consider activity creation via the $2P\to 2A$ channel.
Assuming an activity dynamics with a fractional Laplacian term $|\nabla|^{2\alpha-d}a$ such as \autoref{eq:LR_CDP_a}, one falls back on the results discussed in the previous section.
Instead, we argued in \cite{mariAbsorbingPhaseTransitions2022} that the simple non-linearity $a^2$, leading to the mean-field CDP exponent $\beta_\mathrm{MF}=1$, may be replaced by a term $a^{\gamma}$ with a non-integer
exponent $\gamma$, leading to a modified mean-field exponent $\beta=1/(\gamma-1)$. 
An exponent $\gamma=3/2$, leading to $\beta=2$,
may appear from a simple modelling of the creation of active particles from the diffusion of passive particles in the extreme case of all-to-all interactions ($\alpha=0$)~\cite{mariAbsorbingPhaseTransitions2022}.
This key ingredient related to the small-scale physics of the model may thus result in the emergence of a non-trivial non-linearity of large scale equations.
However, linking physics on sub-particle scales with large-scale physics is a difficult task, and the arguments presented in~\cite{mariAbsorbingPhaseTransitions2022} for $\alpha=0$ are based on a number of simplifications that makes them difficult to extend to account for a spatially-decaying interaction kernel.
In the following, we do not attempt to provide such a connection across scales, but we rather focus on a mean-field description of the creation of activity via diffusion of passive particles, assuming this is the main mechanism responsible for the departure of critical exponents from the short-range CDP values for $\alpha\lesssim d$.

\subsection{Activity creation by passive diffusion: a mean-field approach}

In this section we interpret our numerical results in the light of a mean-field picture 
of activity induced by long-range mediated noise.
This picture is inspired by the one proposed for the yielding transition via which yield stress materials start to flow if subject to a stress $\Sigma$ larger than a yield value $\Sigma_\mathrm{y}$~\cite{bonnYieldStressMaterials2017}.
The yielding transition is analogous to an APT~\cite{jocteurYieldingAbsorbingPhase2024a}: in the flowing phase, flow proceeds by local plastic events analogous to active sites, which induce mechanical load redistribution to elastic parts analogous to passive sites, via a long-range elastic kernel~\cite{nicolas_deformation_2018}. 
The superposition of load redistribution from a finite density of plastic events gives rise to a so-called ``mechanical noise'', which dominates the creation of plasticity.
Remarkably, the yielding transition is also convex: the shear rate $\dot\gamma$, analogous to the average activity, scales as $\dot\gamma \sim (\Sigma - \Sigma_\mathrm{y})^\beta$ with $\beta \approx \numrange{1.5}{2}$~\cite{bonnYieldStressMaterials2017}.
This convex behavior has been rationalized at mean-field level thanks to the Hébraud-Lequeux model (HL), which predicts $\beta=2$~\cite{hebraudModeCouplingTheoryPasty1998}.

In this picture, the statistics of the mediated noise plays a central role. 
We therefore start this section by a detailed mean-field analysis on this noise, before turning to a Hébraud-Lequeux-like description of the transition observed in the mediated ROM, based on the anomalous diffusion that the noise statistics induces.

\subsubsection{Statistics of internal noise}
\label{sec:noise:dist}

We start by determining the probability distribution of the noise received by a given passive particle, i.e., the sum of all the displacements induced by long-range mediated interactions with active particles throughout the system, during a single time step.
To simplify the calculation, we neglect the correlations between the positions of the active particles. This assumption is consistent with other mean-field-type approximations performed in this section.

If there are $n$ active particles at time t, the displacement $\bm{\xi}$ of a given passive particle is expressed as
\begin{equation}
\bm{\xi} = \sum_{j=1}^{n} \bm{\xi}_j
\end{equation}
where $\bm{\xi}_j=(\xi_{x,j},\xi_{y,j})$ is the displacement vector induced by the active particle $j$ situated at a distance $r_j$ from the passive particle.
According to the definition of the mediated ROM, both components $\xi_{x,j}$ and $\xi_{y,j}$ are independent Gaussian random variables with zero mean
and variance $G(r_j) \delta_{a}^2$, where $\delta_{a}^2$ is the variance of the displacement of active particles.
To evaluate the probability distribution of $\bm{\xi}$, we need to average over the distances $r_j$ and over the number $n$ of active particles.
Denoting as $A$ the average fraction of active particles, $A=\langle n\rangle / N$, we assume for $n$ a Poisson distribution of mean $\langle n\rangle=A N$, meaning that each particle is randomly active with probability $A$.
Calculation details are reported in Appendix~\ref{sec:appendix:noise}.
The distribution $P(\bm{\xi})$ obtained in the infinite size limit depends on whether $\alpha<d/2$ or $\alpha>d/2$.
When $\alpha<d/2$, $P(\bm{\xi})$ is a bivariate Gaussian distribution with uncorrelated components $\xi_x$ and $\xi_y$.
The variance of each component is proportional to the activity $A$.
In contrast, for $\alpha>d/2$ the distribution of $\bm{\xi}$ is a broad distribution of parameter $\mu=d/\alpha$,
that is, a distribution with power-law tails
\begin{equation} \label{eq:broad:noise:dist}
    P(\bm{\xi}) \sim \frac{A}{|\bm{\xi}|^{d+\mu}}, \qquad |\bm{\xi}| \to \infty
\end{equation}
(up to constant numerical prefactors).
While the relation $\mu=d/\alpha$ between the exponent $\mu$ of the noise distribution 
and the exponent $\alpha$ of long-range interaction
is known \cite{linMeanFieldDescriptionPlastic2016,linMicroscopicProcessesControlling2018}, and can be straightforwardly obtained by a change of variable argument based on a single active particle,
the calculation presented here goes one step further, as it yields the noise distribution for a given density of active particles in the large system size limit.
As a result, the amplitude of the tails of the distribution $P(\xi)$ is proportional to the density $A$ of active particles [see \autoref{eq:broad:noise:dist}], a result that plays an important role in the following.

\subsubsection{A Hébraud-Lequeux-like model}

\begin{figure}[h]
    \centering
    \includegraphics[width=0.4\textwidth]{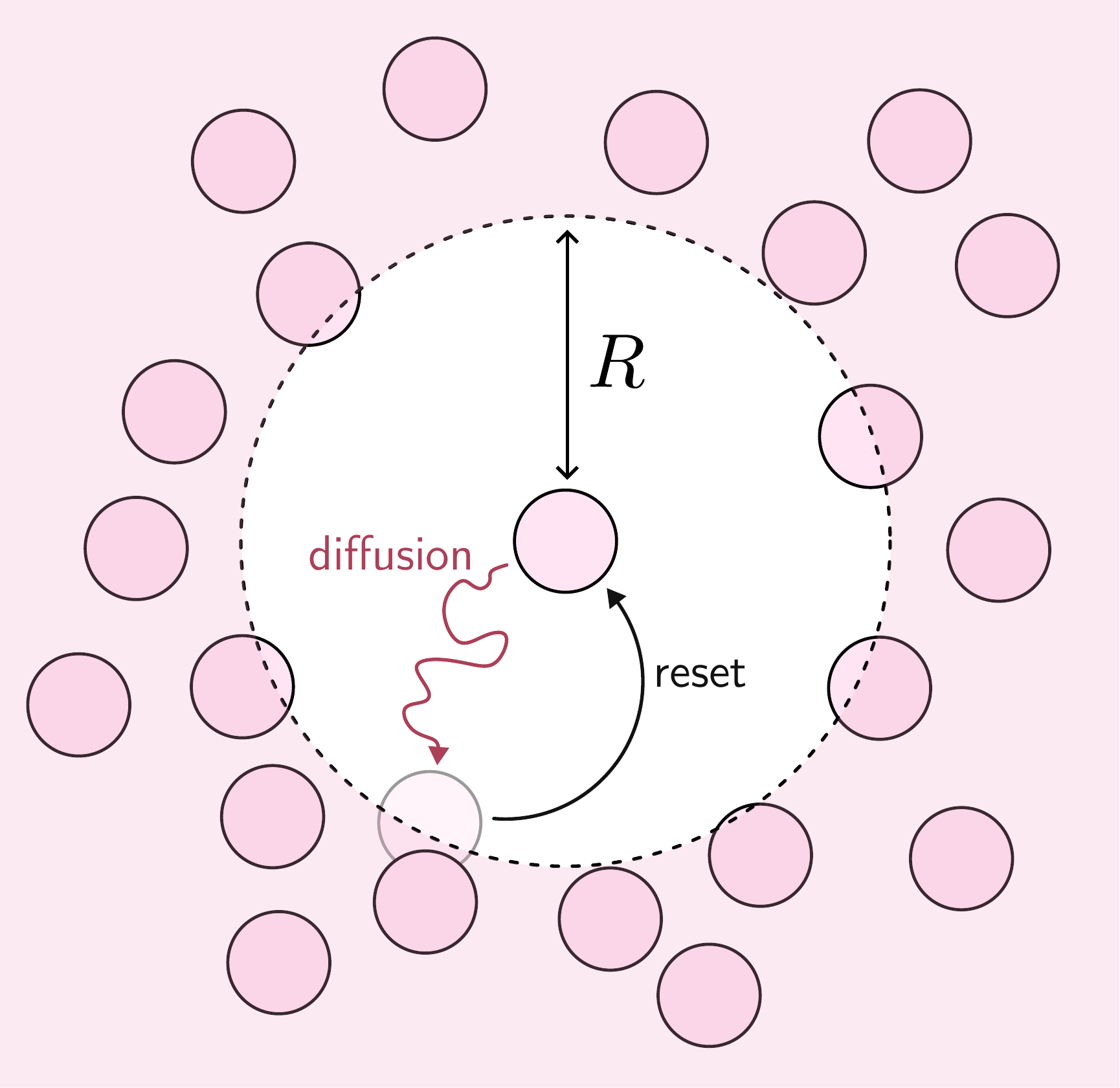}
    \caption{Sketch of the Hébraud-Lequeux-like model for suspensions. A passive particle is surrounded by a locally disordered environment abstracted as an effective cage of radius $R$. The particle diffuses until it reaches the edge of the cage. At this point, the particle goes back to the center of the cage with a given probability rate (i.e., it gets trapped in a new effective cage).}
    \label{fig:LHLmodel}
\end{figure}

Equipped with a mean-field mechanical noise statistics, we now develop an equally mean-field model for the activity creation from two passive particles coming in overlap, that is, $2P\to 2A$.
We consider a single passive particle following a random motion subject to the mechanical noise described in \autoref{eq:broad:noise:dist}. 
This particle is surrounded by neighbors which are also passive and stand at a typical distance $R = (\phi/\Omega_d)^{-1/d}D/2$, with $\Omega_d$ the solid angle in dimension $d$.
The particle becomes active if it encounters one of these passive neighbors, in which case it makes a large jump and lands in a new local environment. 

To model this process at a mean-field level, we trade the local cage made of particles by a fixed effective spherical cage of radius $R$ (see \autoref{fig:LHLmodel}).
When it escapes the cage, the particle becomes active with a rate $\tau^{-1}$, and if so is put back to the center of the cage, to mimic the resetting of the local environment (as seen in the particle frame) once the active particle has become passive again.
The probability distribution $P(\bm{r})$ of finding the particle at $\bm{r}$ then satisfies an evolution equation that depends on the statistics of the noise, \autoref{eq:broad:noise:dist}.
For $\mu\geq 2$, that is, for $\alpha\leq d/2$, the noise distribution is Gaussian, with a variance proportional to the mean activity $A$.
We thus assume that the passive particle is subject to normal diffusion, with a diffusion coefficient proportional to $A$.
We then have 
\begin{equation}
    \partial_t P = \mathcal{D}\nabla^2 P(\bm{r}) - \frac{1}{\tau}\Theta(r>R)P(\bm{r}) + A \delta(\bm{r})\, ,
    \label{eq:HL_1}
\end{equation}
with a regular diffusion term $\mathcal{D}\nabla^2 P$~\cite{hinrichsenNonequilibriumPhaseTransitions2007}, and the mean activity 
\begin{equation}
    A = \frac{1}{\tau} \int_{|\bm{r}|>R}\mathrm{d}\bm{r} P(\bm{r})\, .\label{eq:HL_2}
\end{equation}
The diffusion coefficient $\mathcal{D}$ is proportional to the mean activity 
\begin{equation}
\mathcal{D} = \kappa A\, , \label{eq:HL_3}
\end{equation}
with a parameter $\kappa>0$, which may be determined from the noise distribution.

The model defined by \autoref{eq:HL_1}, \autoref{eq:HL_2} and \autoref{eq:HL_3}, that we dub the \modellong{} (\modelshort) is closely related to the HL model. 
Beyond the obvious difference that our model is $d$-dimensional whereas the usual HL model is one-dimensional (although it is possible to generalize it~\cite{olivierGeneralizationHebraudLequeux2013}), 
the major difference concerns the control parameter. 
The HL model has a driving term involving the deformation rate, which acts as the control parameter, whereas in our model the control parameter is the cage radius $R$, which is directly related to the volume fraction in the mediated ROM as $R\sim \phi^{-1/d}$.

Despite these differences, our model exhibits a phenomenology analogous to the one of the HL model.
The \modelshort{} model displays a phase transition at a critical value $R_\mathrm{c} = \sqrt{2d\kappa}$, as shown in Appendix~\ref{sec:appendix_HL}.
For $R>R_\mathrm{c}$ (corresponding to small volume fractions in the mediated ROM), the steady-state solution of the model is such that $P(\bm{r}) = 0$ if $|\bm{r}|>0$ and consequently $A=0$, 
and we associate this solution with an absorbing state where no activity occurs.
By contrast, for $R<R_\mathrm{c}$ (large volume fractions), the steady-state solution is such that $A>0$.
More specifically, just above the transition we find $A \sim (R_\mathrm{c}-R)^\beta$ with $\beta=2$.

For $\mu<2$, corresponding to $\alpha>d/2$, the long-tailed noise distribution in \autoref{eq:broad:noise:dist} leads to anomalous diffusion, and instead of \autoref{eq:HL_1} we have
\begin{equation}
    \partial_t P = - \mathcal{D}|\nabla|^\mu P(\bm{r}) - \frac{1}{\tau}\Theta(r>R)P(\bm{r}) + A \delta(\bm{r})\, ,
    \label{eq:HL_1_anomalous}
\end{equation}
which defines the \levymodelshort.
Since the tails of the probability distribution of the noise are proportional to the mean activity $A$ 
 (see \autoref{eq:broad:noise:dist}), the generalized diffusion constant $\mathcal{D}$ is still proportional to the activity $A$, so that \autoref{eq:HL_3} remains valid.

\subsubsection{Critical behavior of the \levymodelshort{} model}

\begin{figure}[h]
    \centering
    \includegraphics[width=\linewidth]{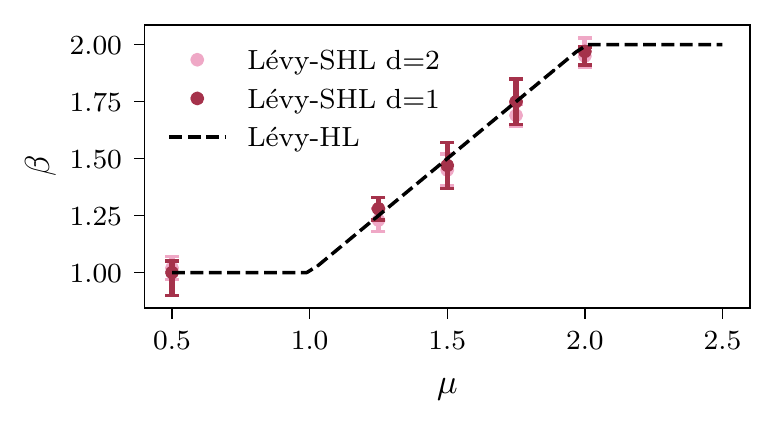}
    \caption{Evolution of the exponent $\beta$ with the Lévy parameter $\mu$ in numerical simulations of the \levymodelshort{} model, \autoref{eq:HL_1_anomalous}, in dimensions $d=1$ and $d=2$.}
    \label{fig:LHL2D}
\end{figure}

Going from \modelshort{} to \levymodelshort{} by replacing normal diffusion by anomalous diffusion in \autoref{eq:HL_1_anomalous} has strong consequences on the criticality, as we now discuss.
We simulate the dynamics of \num{e7} particles following a L\'evy random walk such that their probability distribution follows \autoref{eq:HL_1_anomalous}, and evaluate the exponent $\beta$ as a function 
of $\mu$,  in dimensions $d=1$ and $d=2$. 
Details of the simulations can be found in Appendix~\ref{sec:appendix_LHLnum}.
Remarkably, we find the same relation between $\beta$ and $\mu$ as the one observed (and predicted by scaling theory) for the L\'evy-HL model~\cite{linMicroscopicProcessesControlling2018}, as shown in  \autoref{fig:LHL2D}, 
despite the differences between both models (different control parameters and different dimensions).
For both $d=1$ and $d=2$,  we get $\beta = \mu$ for $1<\mu<2$, while $\beta=1$ for $\mu<1$ and $\beta=2$ for $\mu>2$.

Within the scaling theory for the L\'evy-HL model~\cite{linMeanFieldDescriptionPlastic2016,linMicroscopicProcessesControlling2018}, the $\mu$-dependence of $\beta$ stems from the effect of long-range hoppings on the distribution of the `distance to activity'.
In the context of the yielding transition, the distance to activity is the distance to the local yield stress value: if a mesoscopic element bears a stress $\sigma$ and yields plastically if subject to a stress exceeding $\sigma_\mathrm{c}$, then the quantity $x=\sigma_\mathrm{c} - \sigma$ is the distance to activity.
By contrast, in the present \levymodelshort{} model for suspensions, the distance to activity is $x=R-r$, and its distribution $P_x(x)$ follows a scaling law $P_x(x)\sim x^\theta$ for a vanishingly small activity (similarly to the yielding case~\cite{linMeanFieldDescriptionPlastic2016}), with $\theta$ the pseudo-gap exponent.
For $\mu\geq 2$, for which the behavior is diffusive, $\theta=1$ like in the regular HL model.
For $\mu<2$, $\theta$ depends continuously on $\mu$ and one finds $\theta = \mu/2$ in the absence of driving~\cite{linMeanFieldDescriptionPlastic2016} (when driving is present this is only true for $1<\mu<2$).

As $\mu$ and $\theta$ (contrary to $\beta$) are not related to the order parameter, 
we can anticipate that the relation $\theta(\mu)$ carries over from L\'evy-HL to at least the $d=1$ \levymodelshort.
We have verified the validity of the relation $\theta(\mu)$ in our simulations of the \levymodelshort{} for $d=1$ and $d=2$ (we have not checked it for $d>2$, as running the dynamics gets rapidly prohibitive with increasing dimension).
To accurately measure $\theta$ we use the fact that for small $\epsilon = R_\mathrm{c}-R$ we expect a scaling form $P_x(x) = \epsilon^a f(\epsilon^{-b}x)$ with $a/b = \theta$.
We then look for the values of $a$ and $b$ giving a collapse of the $P_x(x)$ curves for several values of $\epsilon$. Results are shown in \autoref{fig:LHLnum_theta_collapse} of Appendix~\ref{sec:appendix_LHLnum}.
Altogether, this means that within the \levymodelshort, we find $\beta=2$ for $\theta\geq 1$, $\beta=2\theta$ for $1/2 < \theta < 1$, and $\beta=1$ for $\theta<1/2$.

\subsection{Assessing the mean-field picture}

\subsubsection{A stepping stone model}

\begin{figure}[h]
    \centering
    \includegraphics[width=\linewidth]{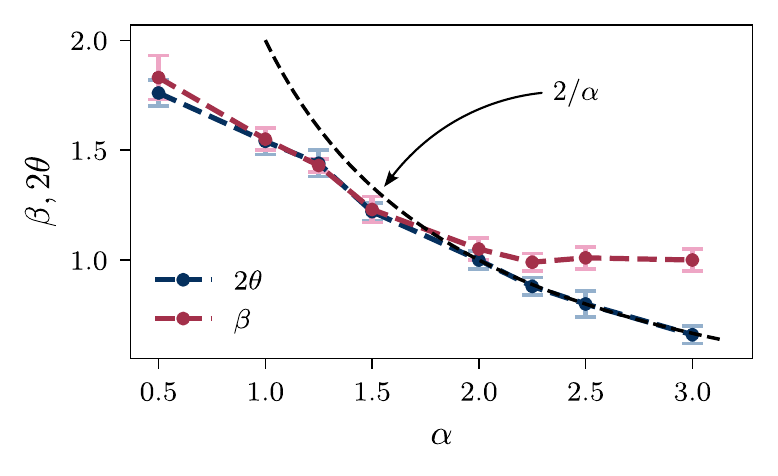}
    \caption{Order parameter exponent $\beta$ and pseudo-gap exponent $\theta$ as functions of the range of interaction $\alpha$ for the \mediatedROMinf{} model with $d=2$.}
    \label{fig:betaMF}
\end{figure}

The \levymodelshort{} model gives us a mean-field framework to describe the $2P\to 2A$ activity creation.
Within the mediated ROM, we expect this channel to dominate for the largest ranges of interaction, that is, in the small $\alpha$ limit. 
For generic values of $\alpha$ however, we expect a competition with the usual CDP-like $A+P\to 2A$ channel, which fully takes over for $\alpha \gtrsim d$.
To assess the mean-field picture of the \levymodelshort{} model, it is therefore convenient to introduce an intermediate model where the $A+P\to 2A$ remains present, but is given a mean-field nature.

This is generically possible in the $\Delta_a\to \infty$ limit. 
More specifically, we consider a model we call \mediatedROMinf{} following the same rules as the mediated ROM, except that when two particles overlap and become active at time step $t$, at time $t+1$ they are moved to new locations picked randomly within a uniform distribution over the system box.
Activity correlations are then mean-field. 

With this model, we can perform direct tests of the mean-field H\'ebraud-Lequeux-like picture given by the \levymodelshort{} model.
We measure the exponent $\beta$ as a function of the range exponent $\alpha$ for the \mediatedROMinf{} with $d=2$, and report the results in \autoref{fig:betaMF} (red curve).
The \mediatedROMinf{} also exhibits a convex transition for small $\alpha$. 
The range of $\alpha$ values over which we observe a floating $\beta$ exponent is the same as for the mediated ROM.
However, in contrast to the mediated ROM, the exponent $\beta$ plateaus at $\beta\approx 1$ in the \mediatedROMinf{} for $\alpha \gtrsim d$.
This is consistent with a regime where now $2P\to 2A$ is irrelevant, and the transition is entirely controlled by $A+P\to 2A$.
As $A+P\to 2A$ has a mean-field nature in the \mediatedROMinf, we indeed expect a mean-field DP behavior with $\beta=1$ in this regime.

In order to numerically measure $\mu$, which is the exponent of the noise distribution in a continuous time dynamics, one needs to pick the relevant time scale over which noise must be evaluated, which is tricky. 
Alternatively one can evaluate $\mu$ by indirect methods measuring the Hurst exponent~\cite{ferreroCriticalityElastoplasticModels2019}.
Here, we rather chose to directly evaluate the exponent $\theta$ in the \mediatedROMinf.
In this model, the distance to activity for a given passive particle is the distance of the closest passive neighbor.
We can then measure $\theta$ using the passive-passive pair correlation function $g(r)$.
The product $\rho_\mathrm{p} g(r)$, with $\rho_\mathrm{p}$ the passive particle density, measures the probability of finding a passive particle at distance $r$ of another passive particle.
By definition $g(r<D)=0$, so we define a gap distribution function $g_x(x)$ with $x=r-D$.
Note that this is not strictly the probability of finding the nearest passive particle at a gap $x$, but for small $x$ we expect that these two functions are asymptotically identical.
We thus expect that at the transition, $\phi=\phi_\mathrm{c}$, we have $g_x(x) \sim x^\theta$, and that slightly away from the transition we have a scaling form $g_x(x) = \delta\phi^a f(\delta\phi^{-b}x)$.
We show in Appendix~\ref{sec:appendix_PCorr} that this expectation is met for all values of $\alpha$, allowing us to measure $\theta=a/b$.

In \autoref{fig:betaMF}, we report the measured values of $2\theta$ as a function of $\alpha$ for the \mediatedROMinf. 
Remarkably, the data perfectly track the measured $\beta$ for $1<2\theta<2$, and markedly deviate from it for $2\theta < 1$, which is exactly the mean-field prediction.
This is a strong evidence that the mean-field picture of the \levymodelshort{} model 
captures the relevant physical mechanisms at work behind the convex transition.

However, the \mediatedROMinf{} is not fully mean-field. 
Recalling the mean-field scaling prediction $2\theta=\mu$, and the mean-field estimate $\mu=d/\alpha$, we get for $d=2$ the mean-field prediction $2\theta = 2/\alpha$ which we also show in \autoref{fig:betaMF} (black dashed line). 
This prediction is in good agreement with our measured values for $2\theta$ for $\alpha\geq 2$, but is a clear overestimate for $\alpha<2$. 
Without measured $\mu$, we can only speculate which of mean-field predictions $2\theta=\mu$ and $\mu=d/\alpha$ is the culprit, if not both.
The fact that in the \mediatedROMinf{} activity correlations are suppressed, in which case $\mu=d/\alpha$ is expected to hold, strongly suggests that $2\theta=\mu$ is violated.
This is consistent with the idea that the mean-field approximation to the passive particle dynamics in the \levymodelshort{} model is not exact due to correlations in the passive particle density field in $d=2$.
If this interpretation is valid, the effect of density correlations is to lower the value of $\theta$ with respect to the mean-field predictions.

We thus have a hybrid situation where the \mediatedROMinf{} is ``partially mean-field'': it follows the mean-field $\beta(\theta)$ but not the mean-field $\theta(\mu)$.
Said otherwise, the finite-dimensional character of the transition is entirely encoded in the non-trivial value of $\theta$.
It is noteworthy that the yielding transition behaves similarly, with the Hurst exponent characterizing the mechanical noise entirely encoding the finite-dimensional character~\cite{ferreroCriticalityElastoplasticModels2019}.

\subsubsection{The mediated ROM in the mean-field light}

\begin{figure}[t]
    \centering
    \includegraphics[width=\linewidth]{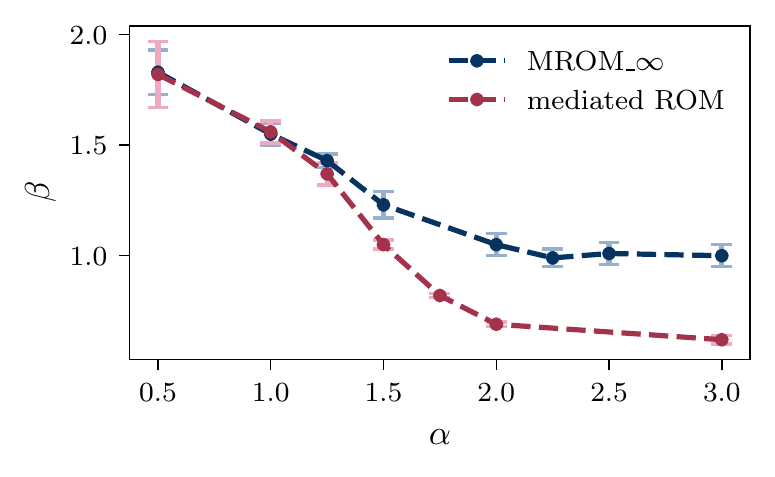}
    \caption{Order parameter exponent $\beta$ as a function of the range of interaction $\alpha$ for the mediated ROM and the \mediatedROMinf model with $d=2$.}
    \label{fig:beta_inf_pasinf}
\end{figure}

\begin{figure}[t]
    \centering
    \includegraphics[width=\linewidth]{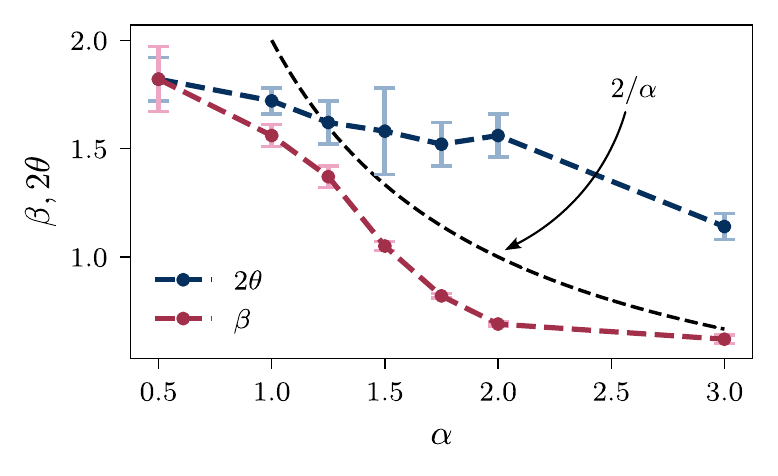}
    \caption{Order parameter exponent $\beta$ and pseudo-gap exponent $\theta$ as functions of the range of interaction $\alpha$ for the mediated ROM with $d=2$.}
    \label{fig:beta_comparison}
\end{figure}

Because of activity correlations, we do not expect any mean-field prediction to be exact in the mediated ROM model.
In \autoref{fig:beta_inf_pasinf}, we compare the measured exponents $\beta$ as a function of the range exponent $\alpha$ obtained for the \mediatedROMinf{} and for the mediated ROM for $d=2$.
For large values of $\alpha$, the value of $\beta$ is systematically larger in the \mediatedROMinf{} than in the mediated ROM. 
The difference is an indirect signature of the role played by $A+P\to 2A$ for short ranges of interactions.
However, decreasing $\alpha$, we observe that the difference shrinks, and for small $\alpha=0.5$, the $\beta$ value is the same within error bars for both models.
This convergence at small $\alpha$ is yet another indirect evidence that in this limit the $2P\to 2A$ channel, which is identical for the two models, is the one controlling the transition, while $A+P\to 2A$ is irrelevant.

This scenario is also supported by the behavior of the pseudo-gap exponent. 
In \autoref{fig:beta_comparison}, we report the measured values for $2\theta$ for the mediated ROM. 
In contrast to the \mediatedROMinf, for large $\alpha$ the value of $2\theta$ is systematically larger than the value of $\beta$.
However, when decreasing $\alpha$,  $2\theta$ and $\beta$ converge to the same value.
Hence, in the small-$\alpha$ limit, not only the $2P\to 2A$ channel governs the transition, but it does so in a partially mean-field way.

\section{Conclusion}

We explored the role of long-range hydrodynamic interactions in the critical behavior of the reversible-irreversible transition observed in suspensions subject to an oscillatory shear, 
thanks to a modified ROM designed to keep the salient features of long-range hydrodynamics within a minimal model.
In particular, hydrodynamics does not impact transport of irreversible activity, 
but rather introduces a mechanism to induce activity over large distance via mechanical noise.
For generic pairwise hydrodynamic mobility falling as $1/r^\alpha$, we find that the RIT turns from concave, that is, with an order parameter exponent $\beta<1$, for large $\alpha$ to convex, $\beta>1$, for small $\alpha$. 
The qualitative change occurs between $\alpha=d-1$ (long-range mobility from bulk hydrodynamics) and $\alpha=d$ (short range mobility typical of confined systems).

Concurrently to this change of convexity, hyperuniformity (suppressed density fluctuations at large length scales), which is a signature of transitions in the CDP class, is gradually washed out when $\alpha$ decreases and disappears completely for convex RITs at small $\alpha$.
We predict that confined systems would show hyperuniformity at the transition, but not bulk systems.
This result imposes to reassess the actual possibility of taking advantage of some RITs to generate hyperuniform states for practical applications~\cite{maHyperuniformityGeneralizedRandom2019}.

Interestingly, while universality classes are well-known to be unstable with respect to long-range interactions, RITs in presence of hydrodynamic interactions do not follow the alterations to conserved directed percolation (CDP, the claimed universality class in absence of hydrodynamic interactions) expected with long-range transport of activity.
Instead, CDP is destabilized towards a criticality controlled by activation from diffusion induced at long range.

We introduced a mean-field picture of diffusion-induced activity, associated with a mean-field model.
This mean-field picture reveals the central role played by the near-contact behavior of the pair correlation function between passive particles, characterized by a critical ``pseudo-gap'' exponent.
It is intriguing that such a short-range feature controls a critical point.
Moreover, the mean-field picture is an analog of the mean-field elasto-plastic scenario of the yielding transition, and indeed our \modelshort{} model is formally very close to the Hébraud-Lequeux model that describes the yielding transition in a mean-field way.
This unexpected and deep connection between the RIT and the yielding transition, which are seemingly quite different, opens avenues for further understanding the RIT in suspensions.
The possibility of a unified description of these two transitions is enthralling.
Finally, this confirms our view that the RIT is a paradigm for a wider class of absorbing phase transitions in soft matter.

It would also be extremely valuable to know how to capture this scenario within a continuum theory. 
While it is clear that the generalization to the CDP continuum theory~\cite{vespignani1998driving,vespignaniAbsorbingstatePhaseTransitions2000a,rossiUniversalityClassAbsorbing2000,menonUniversalityClassReversibleirreversible2009} to include long-range transport will be unable to predict a convex transition, it is far from obvious to us whether an approaching theory could be fitting. 
In~\cite{mariAbsorbingPhaseTransitions2022}, we argued for a non-analyticity in the normal form of the activity dynamics.
However, how the value of the exponent $\alpha$ generically sets this non-analyticity, and even whether the non-analyticity is a necessary or sufficient ingredient for a convex transition, remains to be clarified.
In this direction the RIT, being realized by particle models that are particularly amenable to techniques to derive field theories~\cite{tauberCriticalDynamicsField2013}, could potentially shed light on a putative continuum theory for the yielding transition, which proves quite elusive so far~\cite{jocteurYieldingAbsorbingPhase2024a}.

\acknowledgments
The authors acknowledge fruitful discussions with K. Martens and E.E. Ferrero regarding the characterization of the mechanical noise near the yielding transition.
This project was provided with computer and storage resources by GENCI at
IDRIS thanks to the grant 2023-AD010914551 on the supercomputer Jean Zay's V100 and A100 partitions. 
Some of the computations presented in this paper were performed using the GRICAD infrastructure (\href{https://gricad.univ-grenoble-alpes.fr}{https://gricad.univ-grenoble-alpes.fr}), which is supported by Grenoble research communities.
T.J. acknowledges funding from the French Ministry of Higher Education and Research.


\appendix

\section{Structure factors in the mediated ROM}
\label{sec:appendix:sfact}

\begin{figure}[h]
    \centering
    \includegraphics[width=\linewidth]{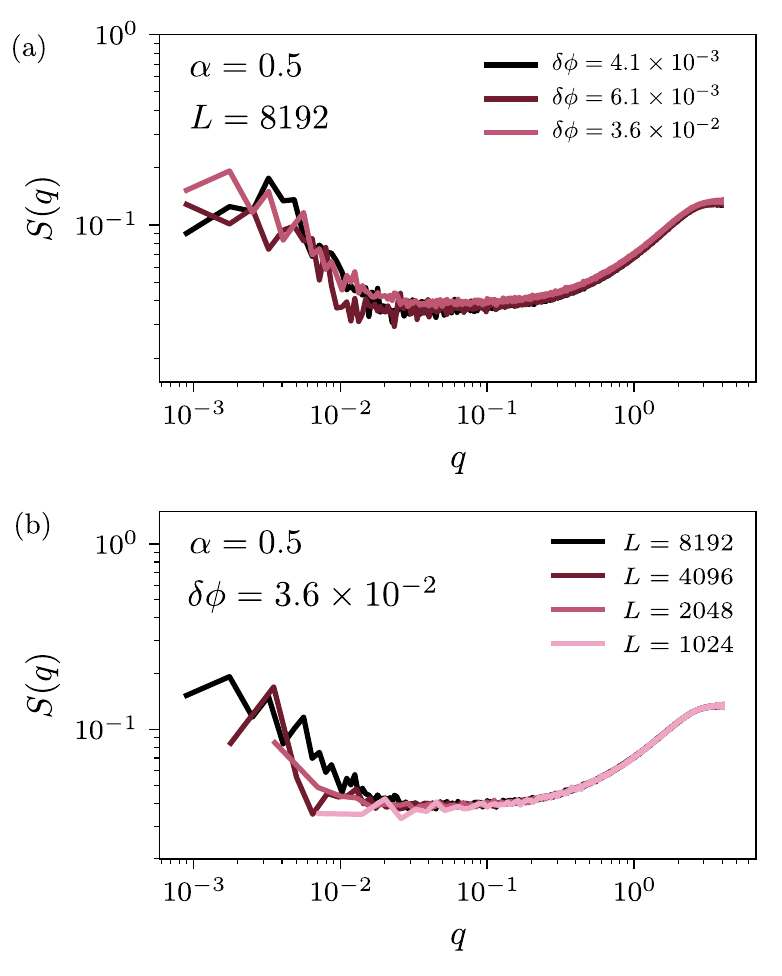}
    \caption{Structure factor in the mediated ROM for $\alpha=0.5$ and $d=2$. (Top) For the largest size $L=8192$, varying the distance $\delta\phi$ to the critical point. (Bottom) For a fixed $\delta \phi$, varying system size.}
    \label{fig:Sq_Appendix}
\end{figure}

We checked numerically the dependence of the structure factor $S(q)$ of the mediated ROM on the distance $\delta\phi$ to the critical point, and on system size, for $\alpha=0.5$. The corresponding structure factors are displayed in \autoref{fig:Sq_Appendix}.
We find no significant dependence of the structure factor on these two parameters, showing that the small-$q$, high plateau value of $S(q)$ does not result from a critical crossover or from finite-size effects.


\section{Evaluation of the noise distribution}
\label{sec:appendix:noise}

In this appendix, we evaluate the noise distribution $P(\bm{\xi})$, where $\bm{\xi}$ is the displacement of a given passive particle during a time step, resulting from long-range mediated interactions
with $n$ active particles in the system: 
\begin{equation} \label{eq:sum:xi:app}
\bm{\xi} = \sum_{j=1}^{n} \bm{\xi}_j,
\end{equation}
with $\bm{\xi}_j$ the displacement induced by active particle $j$, located at position $\bm{r}_j$ with respect to the passive particle situated at $\bm{r}=0$. To simplify the calculation, we assume the $\bm{\xi}_j$'s to be statistically independent. For fixed positions $\bm{r}_j$, the contributions $\bm{\xi}_j$ have a bivariate Gaussian distribution: the components $\xi_{j,x}$ and $\xi_{j,y}$
are statistically independent, have zero mean and variance $G(r) \delta_a^2$, where $\delta_a^2$ is the variance of the displacement of active particles.
To proceed further, it is convenient to work with characteristic functions, that is with the Fourier transform of the probability distribution,
$\chi(\bm{q})=\langle e^{\mathrm{i}\bm{\xi} \cdot \bm{q}} \rangle$, where $\langle \dots \rangle$ indicate an average over $\bm{\xi}$.
For a fixed position $\bm{r}_j$, the characteristic function of $\bm{\xi}_j$ is that of a Gaussian,
\begin{equation} \label{eq:chi1bar:app}
\chi_1(\bm{q},\bm{r}_j) = \exp\left(-\bm{q}^2 g_0^2/2r_j^{2\alpha}\right)
\end{equation}
with $r_j = |\bm{r}_j|$, and approximating the propagator as $G(\bm{r}) \approx g_0^2/r^{2\alpha}$, for $D<r<L$ where $D$ is the particle diameter, and $L$ is the system size.
Averaging over $\bm{r}_j$, we get
\begin{equation}
    \overline{\chi}_1(\bm{q}) = \frac{d}{L^d-D^d} \int_D^L dr \, r^{d-1} \exp\left(-\bm{q}^2 g_0^2/2r_j^{2\alpha}\right).
\end{equation}
Now, for a fixed $n$, the characteristic function $\overline{\chi}_n(\bm{q})$ is given by $\overline{\chi}_n(\bm{q})=\overline{\chi}_1(\bm{q})^n$.
We assume for $n$ a Poisson distribution,
\begin{equation}
    p_n = \frac{\lambda^n}{n!} e^{-\lambda}
\end{equation}
where $\lambda=\langle n \rangle = A N$, with $A$ the fraction of active particles.
Physically, this assumption means that each particle has a probability $A$ to be active, without correlations between particles.
We then find for $\chi_A(\bm{q})=\sum_{n=0}^{\infty} \overline{\chi}_n(\bm{q})$:
\begin{equation} \label{eq:chiAq:app}
    \chi_A(\bm{q}) = \exp\left[-\lambda \left( 1- \overline{\chi}_1(\bm{q})\right) \right].
\end{equation}
For $L\gg D$, we have from \autoref{eq:chi1bar:app}
\begin{equation} \label{eq:chiAq:app2}
    \lambda \left( 1- \overline{\chi}_1(\bm{q})\right) = S_d \rho A \int_D^L dr \, r^{d-1} \left( 1- e^{-q^2 g_0^2/2r^{2\alpha}} \right),
\end{equation}
with $\rho=N/V$ the particle density ($V$ is the volume of the system), and where $S_d=V/L^d$ is a geometrical factor.
We note that the system size $L$ and particle diameter $D$ now appear only as bounds of the integral in \autoref{eq:chiAq:app2}.
For $\alpha>d/2$, this integral converges at both its upper and lower bounds, so that one can formally take the limits $L\to \infty$ and $D\to 0$.
We end up with
\begin{equation}
    \chi_A(\bm{q}) = e^{-c\rho A q^{\mu}}
\end{equation}
with $q=|\bm{q}|$, $\mu=d/\alpha<2$, and $c=2^{-\mu/2} S_d \Gamma(1-\frac{\mu}{2}) g_0^2$, where $\Gamma$ is the Euler gamma function.
The characteristic function $\chi_A(\bm{q})$ is thus very similar to that of a L\'evy distribution, except that it is defined here in dimension $d$.
Taking the inverse Fourier transform, one finds that the noise distribution $P(\bm{\xi})$ has power-law tails,
\begin{equation}
    P(\bm{\xi}) \sim \frac{A}{|\bm{\xi}|^{d+\mu}}, \qquad |\bm{\xi}| \to \infty.
\end{equation}
In the opposite case $\alpha<d/2$, one instead obtains from \autoref{eq:chiAq:app}, for large $L$, a Gaussian characteristic function,
\begin{equation}
    \chi_A(\bm{q}) = e^{-c' \rho A q^2 g_0^2 L^{d-2\alpha}}
\end{equation}
with $c'$ a constant. Since $d-2\alpha>0$, the variance of the noise diverges when $L\to\infty$ if the amplitude $g_0$ of the propagator is kept fixed.
This indicates that $g_0$ has to be rescaled with system size so as to obtain a well-defined limit for $L\to\infty$, as done in the numerical model.


\section{Steady-state solution of the \modelshort{} model}
\label{sec:appendix_HL}

In this section, we compute explicitly the steady-state solution of the model presented in \autoref{eq:HL_1}, \autoref{eq:HL_2} and \autoref{eq:HL_3} in dimension $d$.

\subsection{$d=1$}

For $d=1$, defining dimensionless time $\tilde{t} = t/\tau$ and position $\tilde{x} = \bm{x}/R$, the steady-state distribution satisfies
\begin{align}
    0 &= \tilde{\kappa}\tilde{A}\partial^2_{\tilde{x}} P(\tilde{x}) 
    - \Theta(1-|x|)P(\tilde{x}) + \delta(\tilde{x})\tilde{A}\, , \label{eq:HLlike_1d}\\
    \tilde{A} &= \int_{|\tilde{x}|>1}\mathrm{d}\tilde{x}\ P(\tilde{x})\, ,
    \label{eq:closure_nondim_1d}
\end{align}
with $\tilde{\kappa} = \kappa/R^2$.

As for the usual HL model, we can solve \autoref{eq:HLlike_1d} by splitting the domain and temporarily consider $A$ a constant, and only in a second step require \autoref{eq:closure_nondim_1d} be satisfied~\cite{agoritsasRelevanceDisorderAthermal2015}.
The solution is symmetric with respect to $\tilde{x}=0$, so we can solve only for $0<\tilde{x}<1$ and $\tilde{x}>1$ and enforce continuity of $P$ and $\partial_{\tilde{x}}P$ in $\tilde{x}=1$, to find
\begin{equation}
    P(\tilde{x}) = \begin{cases}
                \frac{c}{\sqrt{\tilde{\kappa}\tilde{A}}} \left[ 1 + \sqrt{\tilde{\kappa}\tilde{A}} - \tilde{x}\right] \quad \text{if $0<\tilde{x}<1$} \\
                c \exp\left[-\frac{\tilde{x}-1}{\sqrt{\tilde{\kappa}\tilde{A}}} \right] \quad \text{if $\tilde{x}>1$}
                \end{cases}
\end{equation}
with $c$ a constant, which is fixed by the normalization of $P$ to
\begin{equation}
    c = \frac{1}{2 + 2\sqrt{\tilde{\kappa}\tilde{A}} + \frac{1}{\sqrt{\tilde{\kappa}\tilde{A}}}}\, .
\end{equation}
The self-consistency relation \autoref{eq:closure_nondim_1d} then leads to
\begin{equation}
    \tilde{A} = \frac{2\tilde{\kappa}\tilde{A}}{1+2\sqrt{\tilde{\kappa}\tilde{A}}+2\tilde{\kappa}\tilde{A}}\, ,
\end{equation}
which we expand for small $\tilde{A}$ to 
\begin{equation}
    \tilde{A} = 2\tilde{\kappa}\tilde{A}\left[1-2\sqrt{\tilde{\kappa}\tilde{A}} +\mathcal{O}(\tilde{\kappa}\tilde{A})\right]\, .
\end{equation}
We then see two cases separated by a threshold value $\tilde{\kappa}_\mathrm{c}=1/2$, much like in the usual HL model~\cite{agoritsasRelevanceDisorderAthermal2015}.
This threshold is to be understood as a threshold $R_\mathrm{c} = \sqrt{2\kappa}$ for the cage radius (and in turn on the volume fraction in the mediated ROM) at fixed $\kappa$. 
For $R>R_\mathrm{c}$ (that is, $\tilde{\kappa}<\tilde{\kappa}_\mathrm{c}$) the only solution is $\tilde{A}=0$.
For $R<R_\mathrm{c}$, there is another solution $\tilde{A}>0$ scaling as
\begin{equation}
    \tilde{A} \sim (R_\mathrm{c} - R)^2\, ,
\end{equation}
so that the transition is convex with $\beta=2$.

\subsection{$d\geq 2$}

The $d=1$ results extend to any dimension, as we show here for $d\geq 2$.
Defining dimensionless time $\tilde{t} = t/\tau$ and position $\tilde{\bm{r}} = \bm{r}/R$, the steady-state distribution satisfies
\begin{align}
    0 &= \tilde{\kappa}\tilde{A}\tilde{\nabla}^2 P(\tilde{\bm{r}}) - \Theta(1-|\tilde{\bm{r}}|)P(\tilde{\bm{r}}) + \delta(\tilde{\bm{r}})\tilde{A} \\
    \tilde{A} &= \int_{|\tilde{\bm{r}}|>1}\mathrm{d}^d\tilde{\bm{r}}~P(\tilde{\bm{r}})\label{eq:closure_nondim}
\end{align} 
with $\tilde{\kappa} = \kappa/R^2$. 
To lighten the notation, in the following we keep dimensionless variables but drop the tilde notation, except for the control parameter $\tilde{\kappa}$.

The steady-state solution is isotropic, so we introduce $f(r) = P(\bm{r})$ with $r=|\bm{r}|$.
We then split the domain in $0<r<1$ (domain I) and $r>1$ (domain II). In domain I, the solution $f_\text{I}$ satisfies
\begin{equation}
    0 = \tilde{\kappa}{A}\frac{1}{r^{d-1}}\partial_r \left(r^{d-1}\partial_r f_\text{I}\right)
\end{equation}
which yields
\begin{equation}
	f_\text{I}(r) = \begin{cases}
	                        -c_1 \ln (r) + c_2 \quad \text{if $d=2$}\, , \\
	                        \frac{c_1}{r^{d-2}} + c_2 \quad \text{if $d>2$}\, ,
                        \end{cases}
\end{equation}
with, $c_1>0$ and  $c_2$ constants.
In domain II, the solution $f_\text{II}$ satisfies
\begin{equation}
    0 = \tilde{\kappa}{A}\frac{1}{r^{d-1}}\partial_r \left(r^{d-1}\partial_r f_\text{II}\right) - f_\text{II}(r)\, ,
\end{equation} 
which yields
\begin{equation}
	f_\text{II}(r) = c_3 \left(\frac{r}{\sqrt{\tilde{\kappa}A}}\right)^{1-d/2}K_{d/2-1}\left( \frac{r}{\sqrt{\tilde{\kappa}A}} \right), \quad c_3 \in \mathbb{R}\, ,
\end{equation}
with $K_{d/2-1}$ the modified Bessel function of the second kind of order $d/2-1$.

From here we treat separately the cases $d=2$ and $d>2$, starting by the latter.
Enforcing continuity of $f(r)$ and its derivative in $r=1$, and normalization, we get for $d>2$ 
\begin{align}
	f_\text{I}(r) &= \begin{multlined}[t]
	    c_3\bigg[\frac{1}{2}K_{d/2-1}\left(\frac{1}{\sqrt{\tilde{\kappa}A}}\right) + m (\tilde{\kappa}A) \\
	    + r^{2-d} \left(\frac{1}{2}K_{d/2-1}\left(\frac{1}{\sqrt{\tilde{\kappa}A}}\right) - m (\tilde{\kappa}A)\right)\bigg]
	\end{multlined}\\
	f_\text{II}(r) &= c_3 r^{1-d/2} K_{d/2-1}\left(\frac{r}{\sqrt{\tilde{\kappa}A}}\right) 
\end{align}
with
\begin{multline}
    \frac{1}{\Omega_d c_3} =  \left(\frac{1}{d} + \frac{1}{2}\right)\left(\frac{K_{d/2-1}\left(\frac{1}{\sqrt{\tilde{\kappa}A}}\right)}{2} + m(\tilde{\kappa}A)\right) \\ 
	- m(\tilde{\kappa}A) + (\tilde{\kappa}A)^{1/2} K_{d/2}\left(\frac{1}{\sqrt{\tilde{\kappa}A}}\right)
\end{multline}
and 
\begin{equation}
m(\tilde{\kappa}A) = \frac{K_{d/2}\left(\frac{1}{\sqrt{\tilde{\kappa}A}}\right) + K_{d/2-2}\left(\frac{1}{\sqrt{\tilde{\kappa}A}}\right)}{(4-2d)\sqrt{\tilde{\kappa}A}}\, .    
\end{equation}
We can then express the self-consistent relation \autoref{eq:closure_nondim}
\begin{equation}
    A = \frac{\sqrt{\tilde{\kappa}A} K_{d/2}\left(\frac{1}{\sqrt{\tilde{\kappa}A}}\right)}{\frac{2+d}{4d}K_{d/2-1}\left(\frac{1}{\sqrt{\tilde{\kappa}A}}\right) + \frac{2-d}{2d}m(\tilde{\kappa}A) +\sqrt{\tilde{\kappa}A}K_{d/2}\left(\frac{1}{\sqrt{\tilde{\kappa}A}}\right)}\, .
\end{equation}
Expanding this relation for small $A$, making use of the asymptotic behavior of Bessel functions for large arguments~\cite{abramowitz1968handbook}
\begin{equation}
    K_\nu(r) \sim \sqrt{\frac{\pi}{2r}} e^{-r} \left(1+\frac{4\nu^2 - 1}{8r} + \mathcal{O}(1/r^2)\right)\, ,
\end{equation}
we get at the lowest orders
\begin{equation}
    A = 2d\tilde{\kappa}A \left[1-2\sqrt{\tilde{\kappa}A} + \mathcal{O}(\tilde{\kappa} A)\right]\, .
    \label{eq:self_consistent_expansion}
\end{equation}

We now briefly show that \autoref{eq:self_consistent_expansion} extends to $d=2$, where continuity conditions at $r=1$ impose
\begin{align}
f_\text{I}(r) & = c_3 \left[ K_0\left( \frac{1}{\sqrt{\tilde{\kappa}A}} \right) -K_1\left( \frac{1}{\sqrt{\tilde{\kappa}A}} \right)\frac{\ln (r)}{\sqrt{\tilde{\kappa}A}}\right]\, , \\
    f_\text{II}(r) & = c_3 K_0\left( \frac{r}{\sqrt{\tilde{\kappa}A}} \right) \, .
\end{align}
Normalization further gives
\begin{equation}
\frac{1}{c_3} = \pi K_0\left( \frac{1}{\sqrt{\tilde{\kappa}A}} \right) + \frac{\pi}{2\sqrt{\tilde{\kappa}A}}K_1\left( \frac{1}{\sqrt{\tilde{\kappa}A}} \right)+2\pi \sqrt{\tilde{\kappa}A}K_1\left( \frac{1}{\sqrt{\tilde{\kappa}A}} \right)
\end{equation}
where we used the property $(r^\nu K_\nu (r))^\prime = -r^\nu K_{\nu-1}(r)$.
Self-consistency then reads
\begin{equation}
	A = \frac{\sqrt{\tilde{\kappa}A}K_1\left( \frac{1}{\sqrt{\tilde{\kappa}A}} \right)}{\frac{1}{2} K_0\left( \frac{1}{\sqrt{\tilde{\kappa}A}} \right) + \frac{1}{4\sqrt{\tilde{\kappa}A}}K_1\left( \frac{1}{\sqrt{\tilde{\kappa}A}} \right)+ \sqrt{\tilde{\kappa}A}K_1\left( \frac{1}{\sqrt{\tilde{\kappa}A}} \right)}
\end{equation}
which can again be expanded at small $A$ to get \autoref{eq:self_consistent_expansion} with $d=2$.

Analyzing \autoref{eq:self_consistent_expansion}, we see that just like for $d=1$ there is a threshold value $\tilde{\kappa}_\mathrm{c}=(2d)^{-1}$, corresponding to a threshold $R_\mathrm{c} = \sqrt{2d\kappa}$ for the cage radius (and in turn on the volume fraction in the mediated ROM) at fixed $\kappa$. 
For $R>R_\mathrm{c}$, the only solution is $A=0$.
For $R<R_\mathrm{c}$, there is a solution $A>0$ scaling as
\begin{equation}
    A \sim (R_\mathrm{c} - R)^2\, ,
\end{equation}
that is, the transition is convex with $\beta=2$.

\section{Numerical resolution of the \levymodelshort{} model}
\label{sec:appendix_LHLnum}

We numerically solve the \levymodelshort{} model, \autoref{eq:HL_2}, \autoref{eq:HL_3} and \autoref{eq:HL_1_anomalous}, by simulating the dynamics of \num{1e7} particles which perform random walks with Lévy flights~\cite{chechkin_fundamentals_2006}.
At each time step, each particle jumps by a vector $\bm{\xi}$, which direction is picked at random (equal probability for left and right in $d=1$, and within a uniform distribution on the unit circle for $d=2$), and which norm 
$\xi= | |\bm{\xi}|$ is picked at random within a Lévy stable distribution $\rho(\xi)$ defined by its characteristic function 
\begin{equation}
	\phi(k) = \int \mathrm{d}\xi \exp(ik\xi) \rho(\xi) = \exp (- |c k|^\mu)\, ,
\end{equation}
with
\begin{equation}
	c = \left( s \Gamma(t) \Delta t \right)^{1/\mu}.
\end{equation}
This ensures that the anomalous diffusion term in the evolution of $P(\bm{r})$ is indeed of the form $|\nabla|^\mu P$~\cite{hinrichsen_non_equilibrium_2007, jespersen_levy_1999}. 
In practice we use the Chambers-Mallows-Stuck method~\cite{chambers_method_1976, weron_chambers_mallows_stuck_1996} to generate Lévy stable distributed random numbers.
Finally, when a particle reaches the region $|\bm{r}|>R$, it can become active with a probability rate $1/\tau$, that is, its position is reset in $\bm{r}=0$.

\begin{figure}[h]
    \centering
    \includegraphics[width=\linewidth]{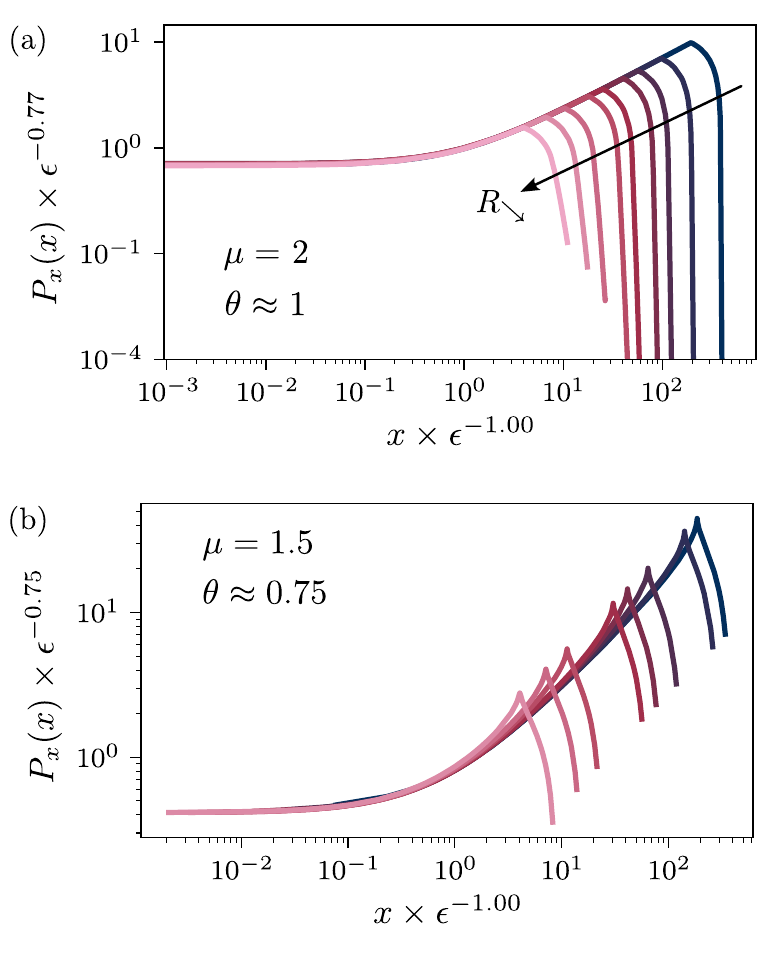}
    \caption{Rescaled gap distribution function $g_x(x)\epsilon^{-a}$ as a function of rescaled gap $x\epsilon^{-b}$ (with $a/b=\theta$) for the \levymodelshort, for noise exponents $\mu=1$ (a) and $\mu=1.5$ (b).}
    \label{fig:LHLnum_theta_collapse}
\end{figure}

In \autoref{fig:LHLnum_theta_collapse}, we show for $d=2$ the gap distribution $P_x(x) = [2\pi(R-x)]^{-1}\int \mathrm{d}\varphi P(R-x)$ for two values of $\mu$, $\mu=1$ and $\mu=1.5$.
In both cases, we rescale the curves as $g_x(x)\epsilon^{-a}$ as a function of $x\epsilon^{-b}$ to obtain a collapse for small values of $\epsilon = R_\mathrm{c}-R>0$, from which we can estimate the pseudo-gap exponent $\theta=a/b$.

\section{Determination of the pseudo-gap exponent in the suspension models}
\label{sec:appendix_PCorr}

\subsection{\mediatedROMinf}

\begin{figure}[h]
    \centering
    \includegraphics[width=\linewidth]{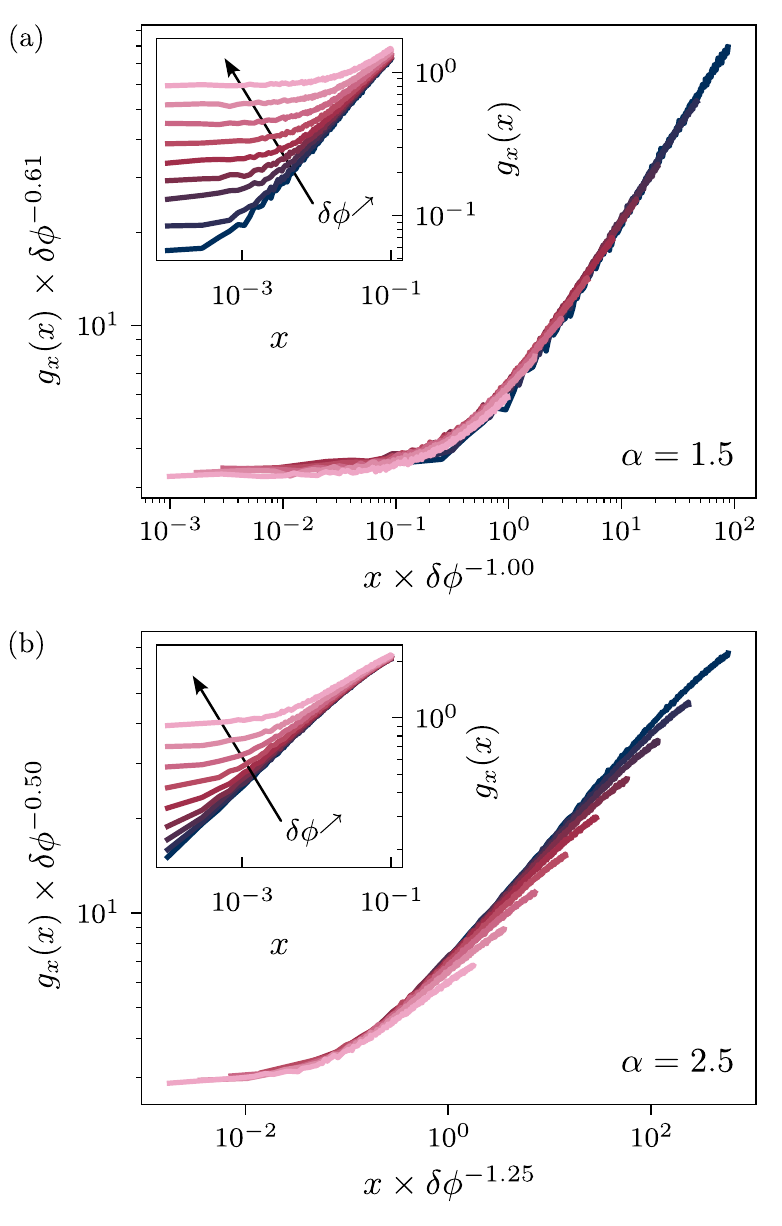}
    \caption{Rescaled gap distribution function $g_x(x)\delta\phi^{-a}$ as a function of rescaled gap $x\delta\phi^{-b}$ (with $a/b=\theta$) for the \mediatedROMinf, for $\alpha=1.5$ (a) and $\alpha=2.5$ (b). Unscaled data in insets.}
    \label{fig:PCorr_alphaMF}
\end{figure}

In order to measure the pseudo-gap exponent $\theta$ in the \mediatedROMinf, we measure the passive-passive pair correlation function $g(r)$ in the stationary state for different distances to the critical point $\delta\phi$ for every range of interaction $\alpha$. Denoting $x = r-D$, we then expect the scaling form $g(x)= \delta\phi^a f(\delta\phi^{-b}x)$ to define the pseudo-gap exponent as $\theta = a/b$. Rescaling all curves according to $x\rightarrow x/\delta\phi^b$ and $g(x)\rightarrow g(x)/\delta\phi^a$ for a given $\alpha$, we look for the couple of values $(a,b)$ which allows for the best collapse onto the same master curve. The error on the measurement of $\theta$ is given by the range of ratios $a/b$ for which the collapse is convincing to the eye. In \autoref{fig:PCorr_alphaMF}, we show the best collapses obtained for $\alpha = 1.5$ and $\alpha = 2.5$. For each $\alpha$ a collapse is possible and the global evolution of $\theta$ is depicted in \autoref{fig:betaMF}.

\subparagraph{}Another information we get from this analysis is that the plateau reached for every $\delta\phi$ at low $x$ scales as $\delta\phi^{\beta/2}$ for any $\alpha$ ($a = \beta /2)$. In other words, at a finite distance from the critical point, the passive-passive pair correlation function takes the form :

\begin{equation}
    g(x) \sim \sqrt{\langle A \rangle} + x^{\theta}
\end{equation}

\noindent which we use for the analysis of the mediated ROM.

\subsection{Mediated ROM}

\begin{figure}[h]
    \centering
    \includegraphics[width=\linewidth]{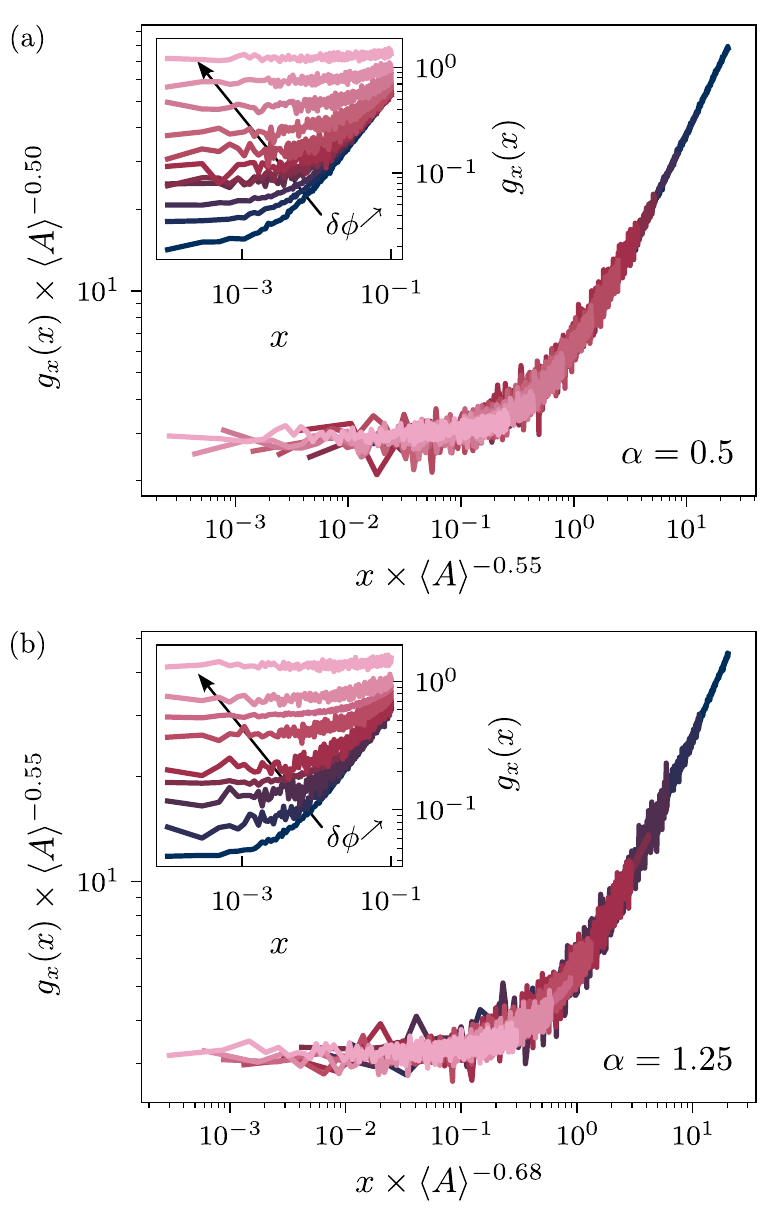}
    \caption{Rescaled gap distribution function $g_x(x)\langle A \rangle^{-a/\beta}$ as a function of rescaled gap $x\langle A \rangle^{-b/\beta}$ (with $a/b=\theta$) for the mediated ROM, for $\alpha=0.5$ (a) and $\alpha=1.25$ (b). Unscaled data in insets.}
    \label{fig:PCorr_alpha}
\end{figure}

To determine $\theta$ in the mediated ROM, that is, with finite jumps of the active particles, we follow the same procedure except for the fact that we now rescale the different curves with $\langle A \rangle$ and not $\delta \phi$. The most convincing collapses obtained for $\alpha = 0.5$ and $\alpha = 1.25$ are shown in \autoref{fig:PCorr_alpha}. Similarly to the \mediatedROMinf, a determination of $\theta$ is possible for every range of interaction and its values are reported in \autoref{table:exponents}.

%

\end{document}